\documentclass[]{aastex631}

\usepackage{bm}
\usepackage{graphicx}
\usepackage{multirow}
\usepackage{soul}
\usepackage{amsmath}
\usepackage{fontawesome}
\usepackage{mathrsfs}
\usepackage{amssymb}
\usepackage{yfonts}
\usepackage{cancel}
\usepackage{color}
\usepackage{xspace}
\usepackage{url}
\usepackage{verbatim}
\usepackage{mathtools}
\usepackage{upgreek}
\usepackage{units}
\usepackage{amstext}
\usepackage{booktabs}
\usepackage{tabulary}
\usepackage{tabularx}
\usepackage{etoolbox}
\usepackage[version=4]{mhchem}
\usepackage[utf8]{inputenc}
\newcolumntype{Y}{>{\centering\arraybackslash}X}
\usepackage[official]{eurosym}
\definecolor{lightgray}{rgb}{0.9,0.9,0.9}	    
\definecolor{green}{rgb}{0,0.5,0}
\definecolor{red}{rgb}{1,0,0}
\definecolor{blue}{rgb}{0,0,0.5}

\usepackage{bbm,bm}
\usepackage{mathrsfs}
\usepackage{slashed}
\usepackage{epstopdf}
\usepackage[normalem]{ulem}
\usepackage[bottom]{footmisc}

\usepackage{bbold}
\usepackage{changepage}
\usepackage[utf8]{inputenc}

\usepackage{grffile}

\usepackage{graphicx}  
\usepackage{bm}        
\usepackage{amssymb}   
\usepackage{amsmath, amssymb}
\usepackage{array}
\usepackage{booktabs}
\usepackage{indentfirst}
\usepackage{float}
\usepackage{lmodern}
\usepackage{multirow}
\usepackage{soul}
\usepackage[normalem]{ulem}

\usepackage{braket}

\newcommand{\dbd}[2]{\ifmmode \frac{\textrm{d}#1}{\textrm{d}#2}\else $\textrm{d}#1/\textrm{d}#2$\fi}
\newcommand{\pbp}[2]{\ifmmode \frac{\partial#1}{\partial#2}\else $\partial#1/\partial#2$\fi}

\DeclareMathAlphabet{\mathpzc}{OT1}{pzc}{m}{it}
 \newcommand{\eV}{\text{e\kern-0.15ex V}\xspace}

 \newcommand{\TeV}{\text{T\kern-0.1ex \eV}\xspace}

\DeclareMathAlphabet{\mathpzc}{OT1}{pzc}{m}{it}

\newcommand{\g}{g_{10}}
\newcommand{\m}{m_a}

\newcommand{\GF}{G_{\rm F}}
\newcommand{\CV}{C_{\rm V}}
\newcommand{\CA}{C_{\rm A}}
\newcommand{\bk}{{\bf k}}

\newcommand{\Mag}{\mathpzc{M}}
\newcommand{\DoBox}[1]{\begin{center}
\color{red}\fbox{
\begin{minipage}{0.9\textwidth}

\end{minipage}}
\end{center}}

\shorttitle{Neutrino Emissivities as a Probe of the Internal Magnetic Fields of White Dwarfs
}

\begin{document}
\title{Neutrino Emissivities as a Probe of the Internal Magnetic Fields of White Dwarfs
}
\reportnum{CP3-21-51}

\author[0000-0003-0521-7586]{Marco Drewes}\affiliation{Centre for Cosmology, Particle Physics and Phenomenology - CP3
Université catholique de Louvain, Chemin du Cyclotron 2
B-1348 Louvain-la-Neuve
Belgium}
\author[0000-0002-9732-8330]{Jamie McDonald}\affiliation{Centre for Cosmology, Particle Physics and Phenomenology - CP3
Université catholique de Louvain
2, Chemin du Cyclotron - Box L7.01.05
B-1348 Louvain-la-Neuve
Belgium}
\author{Lo\"ic Sablon}\affiliation{Centre for Cosmology, Particle Physics and Phenomenology - CP3
Université catholique de Louvain
2, Chemin du Cyclotron - Box L7.01.05
B-1348 Louvain-la-Neuve
Belgium}
\author[0000-0001-7847-1281]{Edoardo Vitagliano} \affiliation{Department of Physics and Astronomy, University of California, Los Angeles\\
475 Portola Plaza, Los Angeles, CA 90095-1547, USA
}

\begin{abstract}

The evolution of white dwarfs (WDs) depends crucially on thermal processes. The plasma in their core can produce neutrinos which escape from the star, thus contributing to the energy loss. While in absence of a magnetic field the main cooling mechanism is plasmon decay at high temperature and photon surface emission at low temperature, a large magnetic field in the core hiding beneath the surface even of ordinary WDs, and undetectable to spectropolarimetric measurements, can potentially leave an imprint in the cooling. In this paper, we revisit the contribution to WD cooling stemming from neutrino pair synchrotron radiation and the effects of the magnetic field on plasmon decay. Our key finding is that even if observations limit the magnetic field strength at the stellar surface,
magnetic fields in the interior of WDs---with or without a surface magnetic field---can be strong enough to modify the cooling rate, with neutrino pair synchrotron emission being the most important contribution.
This effect may not only be relevant for the quantification and interpretation of cooling anomalies, but also suggests that the internal magnetic fields of WDs should be smaller than $\sim 6\times 10^{11}\,\rm G$, slightly improving bounds coming from a stability requirement.
While our simplified treatment of the WD structure implies that further studies are needed to reduce the systematic uncertainties, the estimates based on comparing the emissivities illustrate the potential of neutrino emission as a diagnostic tool to study the interior of WDs.

\end{abstract}



\section{Introduction and review}

Approaching the endpoint of their evolution, stars with masses of up to several solar masses become white dwarfs (WDs). At this stage, the typical mass of the remnant of a once shining star is comparable to the solar mass, $M\sim0.6\,M_{\odot}$, compressed in a radius similar to the Earth one. As in WDs there is no burning nuclear fuel holding up against gravitation, the star would undergo collapse if it were not for the electron degeneracy pressure, as identified in a seminal paper by Fowler~\citep{Fowler:1926zz}. Therefore, while the mass of the star is mostly due to the nuclei, the main contribution to the pressure comes from the electrons. In most of WDs the density is such that electrons are still non-relativistic, and only for rarer large masses one cannot neglect the relativistic corrections to the equation of state, until approaching the Chandrasekhar limit~\citep{Chandrasekhar:1931ih}.

Owing to their high density, WDs can host very strong magnetic fields. The magnetic field at surface of a WD can be inferred from spectropolarimetric measurements.
It turns out 20\% of WDs are known to be magnetic, with fields ranging on the surface from $10^4$ to $10^8$ G (see e.g. \cite{2015A&A...580A.120L,2020AdSpR..66.1025F,liebert2003true,landstreet2019new}). However, the magnetic field in the core of WDs is very poorly constrained and can be much stronger than at surface. For example, \textit{even ordinary} WDs could potentially have internal magnetic fields as large as $10^{12}\,\rm G$~\citep{Shapiro:1983du} or even larger for heavy mass WDs~\citep{Bera:2014wja,Franzon:2015gda}. Possibly, one of the only few ways to constrain the internal magnetic field of WDs is the requirement of the star to be stable through a simple argument firstly introduced in~\cite{chandrasekhar1953problems}.

Besides affecting the structure of the WD,
such strong magnetic fields could modify the cooling of the star, which depends on the local properties. Many of the processes that contribute to the cooling involve the emission of neutrinos, which can escape from regions deep inside the WD. This makes the rate of cooling an observable that is sensitive to the properties of the interior of the WD, which is hidden from direct observation.
In non-magnetic WDs, the cooling proceeds roughly through two different stages.\footnote{In the following, we will use ``non-magnetic WDs'' to indicate WDs with no magnetic field both on the surface and in the core.} When the WD is hot, the main cooling mechanism is through the decay of plasmons, the excitations of the electromagnetic fields in the medium, to neutrinos~\citep{Kantor:2007kf,Winget:2003xf}.\footnote{We call ``plasmon'' any excitation of the electromagnetic field, making no distinction between the polarisation nature of them.} The latter is allowed by the dispersion relations of the photons in a medium, which for transverse excitations in the non-relavistic approximation gives photons a thermal mass~\citep{Adams:1963zzb,zaidi1965emission,Braaten:1993jw,Haft:1993jt}. As the WD gets colder, neutrino emission is suppressed and it cools down through the surface emission of photons. A large magnetic field can potentially catalyze the cooling by modifying the ordinary processes of a non-magnetised core~\citep{DeRaad:1976kd,skobelev1976reaction,galtsov1972photoneutrino,1970Ap&SS...7..407C,Kennett:1999jh,1970Ap&SS...9..453C}, and introducing additional processes as the neutrino pair synchrotron radiation~\citep{Landstreet:1967zz,1981AN....302..167I,Kaminker:1992su}---i.e., the emission of neutrinos by electrons scattering on external magnetic fields. We will show that the largest impact of the magnetic field on cooling is due to synchrotron radiation.

In this paper we examine the impact of 
magnetic fields in the interior of WDs 
on their cooling rates, which is limited both at the population level via the white dwarf luminosity function (WDLF) and at the level of individual stars via pulsation measurements. 
Our main goal is to explore the possibility to impose upper bounds on the magnitude of the magnetic fields. 
In addition, we comment on the possibility that magnetic fields can account for several anomalies in the observed cooling of WDs.
We summarise the effects of magnetic fields on WD cooling in Fig.~\ref{fig:all_processes}.

Constraining the 
magnetic field in degenerate stars complements several recent observations. Astrometric measurements suggest that most of red giants (RGs), i.e. stars with compact energy sources at the center and a large convective envelope (e.g. with a large convective envelope surrounding a thin hydrogen-burning shell and a degenerate helium core), could host very large magnetic fields~\citep{Fuller_2015,Stello_2016,2016ApJ...824...14C}, with these magnetic fields buried under the He raining down from the H burning shell. Any magnetic field would then stay hidden below the envelope due to its long diffusion Ohmic timescale. Similarly WDs could also hide large magnetic fields beneath their surface~\citep{2016ApJ...824...14C}.

From a fundamental physics point of view, there has been an ever growing interest in the use of stars  as laboratories of particle physics~\citep{Raffelt:1996wa}, e.g.~in the context of feebly interacting particles \cite{Agrawal:2021dbo}.
In the realm of weakly interacting slim particles (WISPs)~\citep{Arias:2012az}, astrophysical environments like RGs and WDs have helped defining constraints in the parameter space of these new particles (see~\cite{Raffelt:1996wa,DiLuzio:2020wdo} for reviews). WISPs may also account for the total fraction of observed dark matter, when produced e.g.~through the misalignment mechanism~\citep{Preskill:1982cy,Dine:1982ah,Abbott:1982af} and potentially solve other problems plaguing the Standard Model (SM) of particle physics (e.g. the QCD axion, defined by its coupling to gluons, can explain the smallness of CP-violation in the strong sector, solving the so-called strong CP problem~\citep{Peccei:1977hh,Dine:1981rt}).\footnote{Note that it has recently been put into question in \cite{Ai:2020ptm} whether there actually is a strong CP problem in the SM.}
Crucially, several astrophysical systems show a mild preference for an additional cooling channel~\citep{Giannotti:2015kwo,Giannotti:2017hny}. The bulk of these observations constitutes the so-called ``star cooling excess'', and it has been advocated to be a hint of existing WISPs (see e.g.~\cite{Isern:1992gia} for an early speculation). However, the stellar cooling excess is a discrepancy between the predicted and observed energy losses. As claims of physics beyond the SM should require scrutiny of SM solutions, one has to check for any possible enhancement of the predicted energy loss due to overlooked effects, e.g.~a large magnetic field hiding beneath the surface of degenerate stars.

The present study is also motivated by the rapidly improving situation on the observational side.
As WDs have extinguished all the nuclear fuel in their core, their temperature evolution is determined solely by cooling. WDs can be studied through different observables: both single stars and the WDLF 
see e.g.~\cite{1988ApJ...332..891L}, can be used to study their cooling. The advent of cosmological surveys like the Sloan Digital Sky Survey (SDSS) (see e.g.~\cite{Harris_2006,2017AJ....153...10M}) and the Super COSMOS Sky Survey~\citep{10.1111/j.1365-2966.2011.18976.x} greatly improved the precision of the luminosity function as the sample size increased to several thousands of stars.

The paper is structured as follows. In Section~\ref{noBcooling} we briefly summarize the cooling of WDs obtained assuming a non-magnetised plasma in the core. In Section~\ref{Sec:large} we revisit the idea of a large magnetic field in the core. Then, we will show how the magnetic field implies both the modification of plasmon decay, described in Section~\ref{Sec:plasmonmagnetic}, and the possibility for electrons to radiate neutrinos through synchrotron radiation, discussed in Section~\ref{Sec:NSR}. This will bring us to our conclusions.

\section{Cooling in absence of magnetic fields}\label{noBcooling}

Before investigating the effect of magnetic fields on WD cooling, 
and in order to put our results into context,
we briefly review the most important cooling mechanisms that do not rely on magnetic fields.

\subsection{Cooling through plasmon processes}
\label{Sec:non-magnetised}

\begin{figure}[ht!]
\plottwo{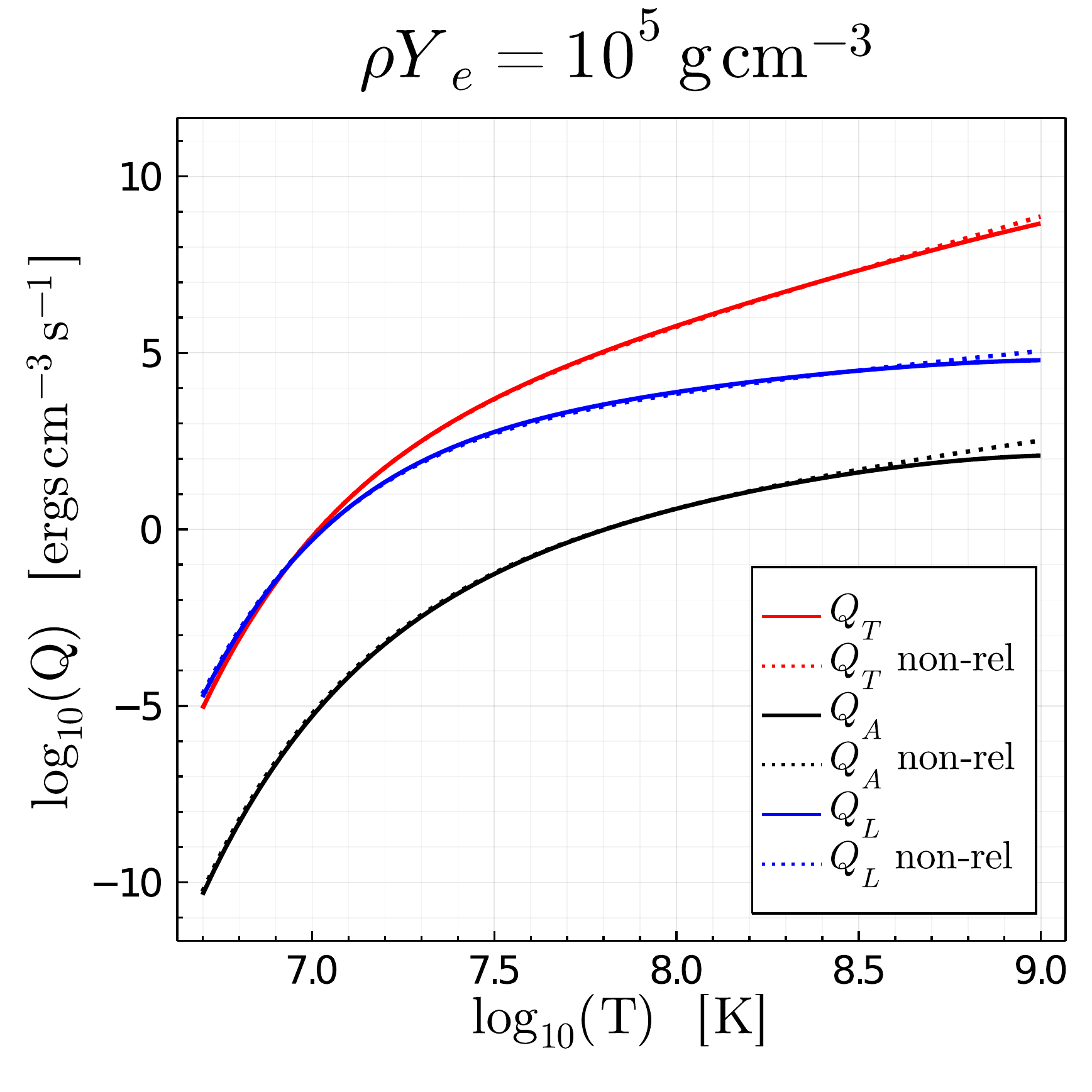}{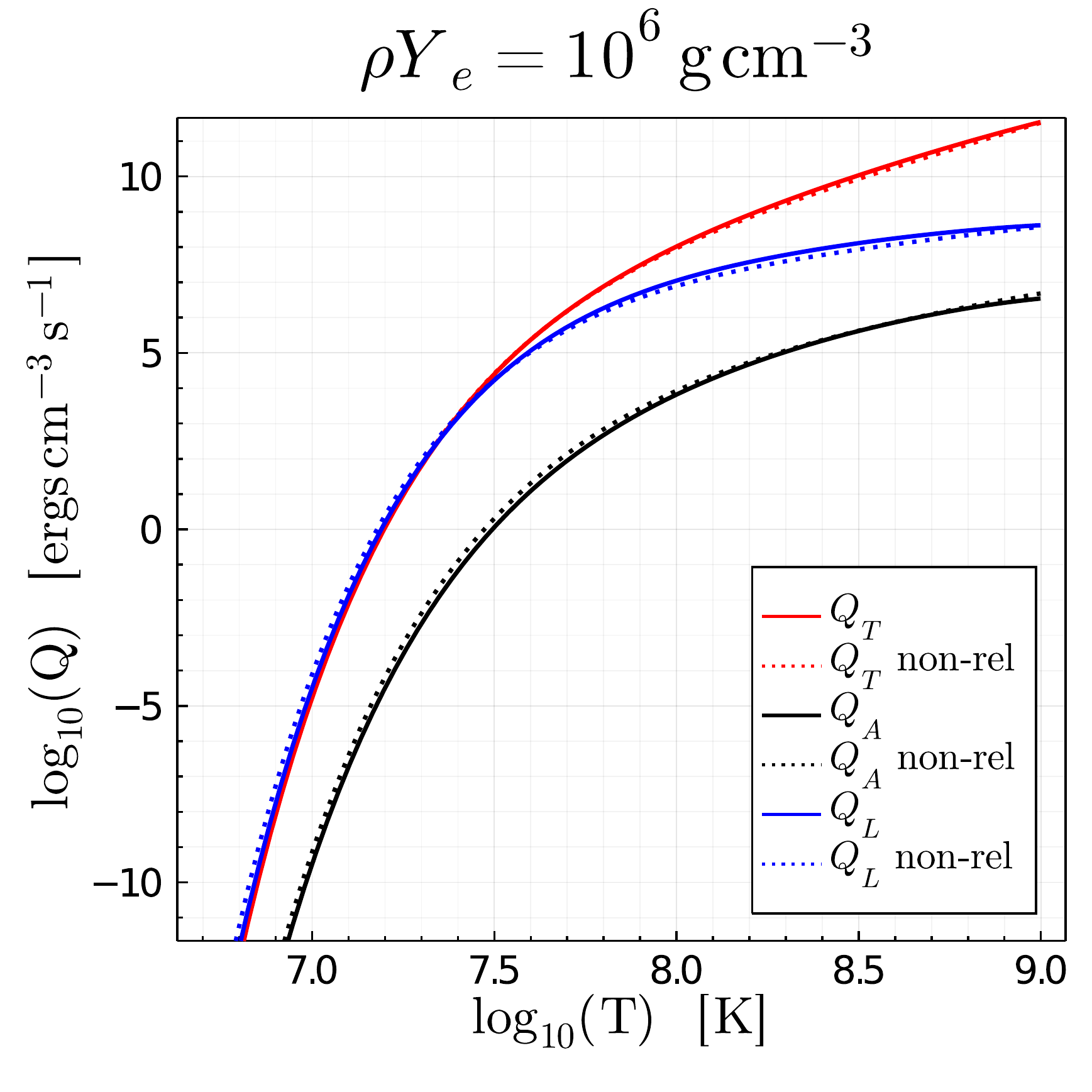}
\plottwo{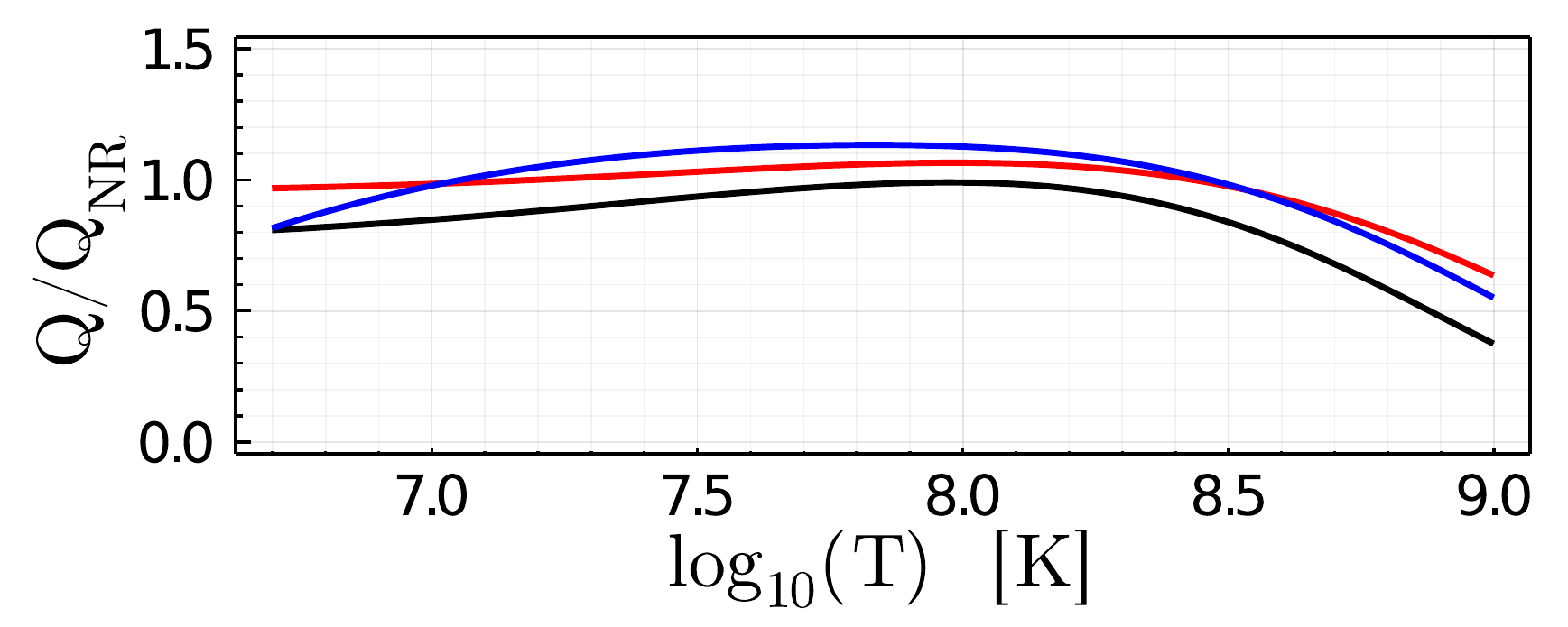}{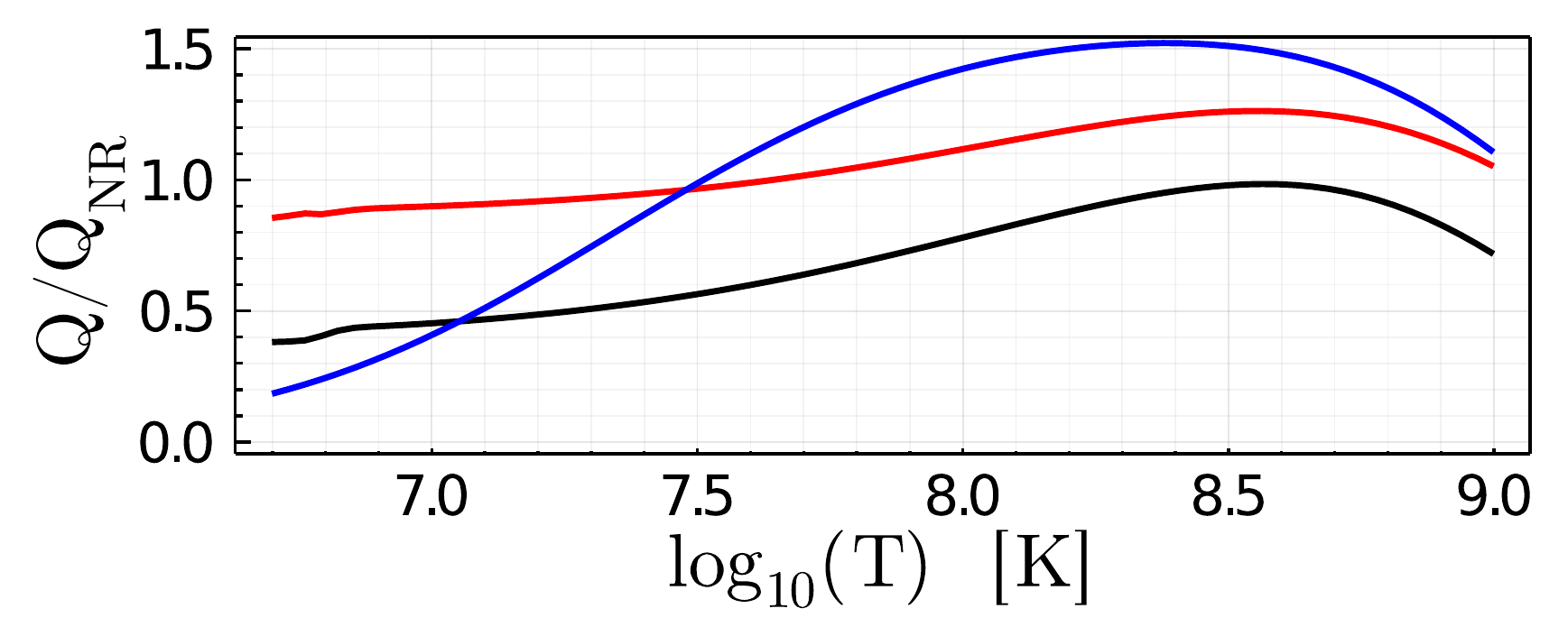}
  \caption{\textbf{Plasmon decay emissivities.} The emissivities in an unmagnetised medium for plasmon decay for the transverse ($Q_T$), longitudinal ($Q_L$) and axial ($Q_A$) processes. We show both the the full emissivities Eqs.~\eqref{eq:QTFUll}--\eqref{eq:QLFUll} (solid lines) and the non-relativistic limits (dotted lines) for $Q_T$ and $Q_A$ given by the integrals in Eqs.~\eqref{eq:qt},\eqref{eq:qaNR} respectively. The bottom panel shows the ratio of the full to non-relativistic emissivities.}
    \label{fig:plasmon}
\end{figure}

The hottest and brightest WDs correspond to the younger ones. During this stage, the dwarf cools mostly through neutrino emission. The dominant emission is from plasmon decay~\citep{Kantor:2007kf}, with other processes like bremsstrahlung coming far second~\citep{2004ApJ...602L.109W}. Plasmon decay in a non-magnetised medium has been extensively studied in the literature~\citep{Adams:1963zzb,zaidi1965emission}, with the most comprehensive treatment given by the seminal paper \cite{Braaten:1993jw} which we use as the basis for our calculations. 
Explicitly, for the energy regimes relevant for stellar cooling, plasmon processes proceed via Fermi interactions,
\begin{equation}\label{eq:NC-interaction}
  {\mathcal L}_{\rm int}=\frac{\GF}{\sqrt{2}}\,
  \bar e\gamma^\mu(\CV-\CA\gamma_5 ) e\,
  \bar\nu_a\gamma_\mu(1-\gamma_5)\nu_a\,,
\end{equation}
where $\GF=1.166\times 10^{-5}\,\rm GeV^{-2}$ is Fermi's constant and $e$ and $\nu_a$ are spinors representing electrons and neutrinos of flavour $a=e, \mu, \tau$, respectively. The effective vector (V) and axial-vector (A)
coupling constants include both neutral current and charged current. Altogether
one finds
\begin{subequations}
\begin{align}
 &\mbox{$\CV=\frac{1}{2}(4\sin\Theta_{\rm W}+1)$
    \qquad and\quad$\CA=+\frac{1}{2}$ \qquad for $\nu_e$,}\\
   &\mbox{$\CV=\frac{1}{2}(4\sin\Theta_{\rm W}-1)$
    \qquad and\quad$\CA=-\frac{1}{2}$ \qquad for $\nu_\mu$, $\nu_\tau$,}
\end{align}
\end{subequations}
in terms of the weak mixing
angle $\Theta_{\rm W}$. One can obtain the squared matrix element for the transition $\gamma\rightarrow \bar{\nu}\nu$ and compute the decay rate
\begin{align}\label{eq:Gamma-plasmon}
  \Gamma_{\gamma\to\nu\bar\nu}\,(\omega)=
  \int\frac{d^3\bk_{\nu}}{(2\pi)^3 2\omega_{\nu}}\,\frac{d^3\bk_{\bar{\nu}}}{(2\pi)^3 2\omega_{\bar{\nu}}}\frac{|{\cal M}_{\gamma\to\nu\bar\nu}|^2}{2\omega}\,
  (2\pi)^4\,\delta^4(k-k_\nu - k_{\bar{\nu}} )\,
\end{align}
where $k_{\nu, \bar{\nu}}=(\omega_{\nu, \bar{\nu}},\mathbf{k}_{\nu, \bar{\nu}})$ and $(\omega,\textbf{k})$ are the wave vectors of outgoing neutrinos and incoming photon, and $\mathcal{M}$ the relevant scattering amplitude. %
The energy loss (per volume per unit time) for each polarisation can be computed as
\begin{equation}
  Q_{\gamma\to\nu\bar\nu}=\int\frac{d^3\bk}{(2\pi)^3}\omega\Gamma_{\gamma\to\nu\bar\nu}(\omega)f_B(\omega) ,
\end{equation}
where $f_B(\omega)=1/(e^{\omega/T}-1)$ is the Bose-Einstein distribution evaluated at the plasmon energy $\omega=\omega_t, \omega_l$ given in Appendix \ref{App:plasmon}. Integrating the energy loss over the star volume, one obtains the luminosity. We can identify three contributions to the plasmon emissivity $Q_{\gamma\to\nu\bar\nu}$ which depend on the polarisation of the photon and whether the process is due to the axial-vector or the vector coupling. These are, calling the fine structure constant~$\alpha=e^2/4\pi$,
\begin{subequations}
	\begin{align}
	\begin{split}
		Q_T = 2\left(\sum_{\nu_\alpha} C_{V}^{2}\right) \frac{G_F^2}{96\pi^4 \alpha} \int_0^{+\infty} dk\, k^2  Z_t(k) \left(\omega_t^2 - k^2\right)^3 f_B(\omega_t)
		\label{eq:QTFUll},
\end{split}
\\
\begin{split}
		Q_A  = 2\left(\sum_{\nu_\alpha} C_{A}^{2}\right) \frac{G_F^2}{96\pi^4 \alpha} \int_0^{+\infty} dk\, k^2 \Pi_A(\omega_t, k)^2 Z_t(k) \left(\omega_t^2 - k^2\right)f_B(\omega_t) ,
 \label{eq:QAFUll}
 \end{split}\\
 \begin{split}
		Q_L  =\, \left(\sum_{\nu_\alpha} C_{V}^{2}\right) \frac{G_F^2}{96\pi^4 \alpha} \int_0^{k_\text{max}} dk\, k^2   Z_l(k) \omega_l^2 \left(\omega_l^2 - k^2\right)^2 f_B(\omega_l) ,	\label{eq:QLFUll}
		 \end{split}
\end{align}
\end{subequations}
which we refer to respectively as transverse, axial and longitudinal. Full expressions for  $Z_\ell$, $\omega_\ell$, $\omega_t$ and $\Pi_A$ can be found in Appendix \ref{App:plasmon}. We display the plasmon emissivities\footnote{Notice that Figure 2 of~\cite{Braaten:1993jw} seems to report the incorrect axial-vector flux, as observed in~\cite{1994ApJ...431..761K}.} Eq.~\eqref{eq:QTFUll}-\eqref{eq:QLFUll} in Fig.~\ref{fig:plasmon}. Note in particular the sub-dominance of the axial contribution.
In the remainder of this paper, we make frequent use of non-relativistic approximations characterised by  $p \ll m_e $  and $T \ll  m_e$, where $p$ is a typical electron momentum and $m_e$ is the electron mass. 
For consistency we examine in Appendix \ref{nonrelaplasmon}
 to what extent the relativistic results for plasmon cooling Eqs.~\eqref{eq:QTFUll}--\eqref{eq:QLFUll} are well-approximated by their non-relativistic limits for the parametric regimes relevant for WDs, thereby providing a cross-check that relativistic corrections can be reasonably neglected in the remainder of the paper. 

\subsection{Cooling through photon surface emission}

At colder temperatures, WDs cool predominantly through photon-surface emission.
This can be estimated by Mestel's cooling law~\citep{mestel1952theory,kaplan1950cooling,Shapiro:1983du}. As the interior of the star is degenerate, electrons have a long mean free path. Therefore, there is a roughly homogeneous temperature all over the core (see for example Figure 1 of~\cite{BischoffKim:2007ve} and Figure 4 of~\cite{Corsico:2014mpa}). On the other hand, the external layers are non-degenerate and are in radiative equilibrium. Thus, one can show that for typical values of a WD, the emission from the non-degenerate external layer is approximately determined by the temperature $T$ in the transition region between the degenerate core and the non-degenerate external layers, approximately equal to the central temperature~\citep{Shapiro:1983du},
\begin{equation}\label{eq:mestellum}
    L_\gamma\simeq (2\times 10^6 \,\mathrm{erg}\, \mathrm{s}^{-1})\frac{M}{M_\odot}T^{3.5} ,
\end{equation}
where $M$ is the WD mass.
 As noted by Mestel, this can be understood as the WD radiating away its residual ion thermal energy.\footnote{Notice that for magnetic fields $B\lesssim 10^{12}\, \rm G$, one can assume the effect of the magnetic field on the specific heat to be negligible (see e.g.~\citep{Baiko:2009rk,Bhattacharya:2015rpa}).}  This is a surface emission, and therefore depends on the global properties of the star. For comparison with other rates given per unit volume, it is useful to define the following volume-averaged emissivity, obtained dividing left and right terms of Eq.~\eqref{eq:mestellum} by the volume,
\begin{equation}\label{VolumeAveragedQgamma}
		 Q_\gamma = \vartheta\frac{L_{\odot}}{M_\odot}  \rho T_{7}^{3.5}
	\end{equation}
where $L_\odot = 3.83 \times 10^{33}\, \mathrm{erg}\,  \mathrm{s}^{-1}$, $\vartheta = 1.7 \times 10^{-3}$, and $T_7=T/10^7\, \rm K$~\citep{Raffelt:1985nj}.\footnote{Notice that the surface emission used in~\cite{Giannotti:2015kwo} is slightly smaller, so we conservatively assume a larger photon emission.}

After this stage, the core crystallizes, and the WD becomes fainter and fainter. The crystallization phase has been recently observed~\citep{tremblay} by the Gaia satellite~\citep{Prusti:2016bjo}, which could improve the accuracy of methods used to determine the age of stellar populations.

\section{Magnetic field effects on white dwarf cooling}
Having reviewed the standard cooling microphysics of a non-magnetic WDs via neutrinos and photons, we now go on and study two new contributions to WD cooling: modifications to the axial emission due to polarisation of the medium by a magnetic field, and neutrino emission via synchrotron processes. The final results are summarised in Fig.~\ref{fig:all_processes}.

\subsection{Revisiting large magnetic fields in WD cores}
\label{Sec:large}

The internal magnetic field $B$ in WDs is poorly constrained by observations and could be much stronger than the one at the surface. The existence of a $B$ field on the surface of magnetic white dwarfs (MWDs) is observed via spectroscopic measurements~\citep{2015A&A...580A.120L,2020AdSpR..66.1025F,liebert2003true,landstreet2019new}. The presence of a large magnetic field, up to hundreds of MG, on the surface of MWDs was observed already in the 1970s by detection of broadband circular and linear polarisation~\citep{1970ApJ...161L..77K,1971ApJ...164L..15A}. Another way of detecting such huge fields is by relying on the Zeeman effect, which can cause the splitting of Balmer lines~\citep{1970ApJ...160L.147A}.

In this way, MWDs have been discovered with $B$ fields of $10^3-10^9 \, \rm G$, in number ever growing, from the circa 70 of the early 2000s~\citep{2000PASP..112..873W} to 600 and counting~\citep{Ferrario2015}, thanks to the Sloan Digital Sky Survey~\citep{2013MNRAS.429.2934K}. The data suggest that $10-20\%$ of WDs show a surface magnetic field (see e.g.~\cite{2014AJ....147..129S,2020AdSpR..66.1025F}). However, we cannot constrain the $B$ field in the core of WDs by spectropolarimetric observations, thus making viable the possibility of a very large magnetic field in the core, possibly even of WDs which do not appear magnetic at the surface. The $B$ field in the core of a WD could be as large as $10^{13}\,\rm G$~\citep{Angel:1978wu,Shapiro:1983du}. Interestingly, MWDs were in the past considered to be WDs which got rid of the Hydrogen envelope, and it was suspected that all WDs had large $B$ fields~\citep{imoto1971evolutionary,1972A&A....16..149C}. However, this hypothesis has been abandoned since magnetic stars have been observed with H dominated atmospheres.  Nevertheless, it might still be the case that even WDs showing no magnetic field on their surface might host large magnetic field in their core.

The problem of measuring the internal magnetic field arises for stars going through different stages of their lives. For example, the sun may host a $B$ field as large as $10^7 \,\rm G$, despite having a much smaller field on the surface~\citep{2003ApJ...599.1434C}. In the case of the sun, bounds can come from neutrinos (as they track the temperature, and the latter would be affected by a different energy density due to the presence of a $B$ field), and from the oblateness of the sun itself~\citep{2003ApJ...599.1434C,2004ApJ...601..570F}. Helioseismology can only put bounds on the magnetic field down to the
tachocline~\citep{2009ApJ...705.1704B,2000A&A...360..335A,2017A&A...601A..47B}. 
Concerning the core of RGs, 
a progenitor of WDs,
the presence of strong magnetic fields (in excess of $10^5 \, \rm G$ or more) is inferred using asteroseismology.
In~\cite{Fuller_2015} it is shown how certain oscillations (so-called g-dominated mixed modes) 
observed at the surface of RGs are damped by the presence of a strong internal magnetic field above a critical value. This critical value turns out to be $10^5\, \rm G$. In a sample of 3000 RGs~\citep{Stello_2016}, about 20\% shows the presence of suppressed dipole modes. Assuming that this reveals the presence of magnetism, 20\% of RGs might possess strong internal magnetic fields. Notice that these fields are likely confined to the stellar core, and do not extend to the stellar envelope/stellar surface. 

Interestingly, asteroseismological observations of RGs suggests that a large fraction of lower mass WDs might have strong internal magnetic fields~\citep{2016ApJ...824...14C}. The key observation is that their magnetic fields are likely buried below the surface and not detectable via Zeeman spectropolarimetry. A few caveats are nevertheless in order. Even if strong B-fields are generated by convective cores during the main sequence, it is not clear if these fields are preserved all the way to the WD formation. There are many processes that could potentially destroy the B-field as the star goes e.g. through the He-flash, or other later convective phases. Moreover, a problematic issue is the stability of the magnetic field itself. Both purely poloidal and toroidal fields are dynamically unstable, even though a combination of the two could be stable~\citep{2010ApJ...724L..34D,2006A&A...450.1077B}. This field can hide under the surface, or emerge at the surface, as long as it has both a poloidal and toroidal component. As the problem of the magnetic field stability is well beyond the scope of this paper, we will neglect these issues.

A final comment should be about variable stars. There is the possibility that pulsating WDs might host a large $B$ field underneath its surface, and hidden even to asteroseismology.  An order of magnitude estimate is enough to show that an extremely large $B$ field would affect the amplitude of oscillations~\citep{Cox1980}. Moreover, if $B$ fields affect oscillations, even a small variation over time of the $B$ field would be enormously amplified by the oscillations. The possibility of using asteroseismology to detect the $B$ field in pulsating WD has been questioned~\citep{Heyl:2000ky,2018MNRAS.477.5338L}. However, we stress that very large magnetic fields should be potentially bounded by asteroseismology in pulsating WDs (see also \cite{1989ApJ...336..403J}).

With these points in mind, the question left is to develop some understanding of the internal magnetic field structure of WDs. In Appendix \ref{App:stability} we use virial arguments to examine an upper bound (see Eq.~\eqref{DeathMagnetic}) on the maximum strength of $B$ that can be supported within some core region whilst maintaining stellar stability. 
The strength of these arguments lies on their simplicity---as they are build on very basic physical principles, and thus in principle very robust. However, by construction virial arguments can only constrain the integrated magnetic field within some region. Applying them to the star as a whole yields a rather robust upper limit on the average value of $B$, which could, however, easily be avoided by much stronger fields that are localised in small regions inside the WD. Applying virial arguments to sub-regions of the WD relies on a detailed understanding of its structure, leading to a loss of the robustness due to modelling uncertainties.
In the remainder of this section we present an alternative way to constrain the local $B$ in sub-regions of the WD, based on its impact on the neutrino emissivity.

\subsection{Plasmon decay in a magnetised medium}

\label{Sec:plasmonmagnetic}

Neglecting processes which depend on the ion content (such as crystallization), a non-magnetised QED plasma is completely described by the electron number density (or alternatively the chemical potential) and the temperature. The presence of a macroscopic magnetic field, as we will review, can affect the plasma in several ways. 
Firstly, it can generate QED nonlinear effects. Moreover, it can affect the electron wave-functions, and modify the propagation of the electromagnetic field excitations. Some of these effects will modify the energy loss from plasmon decay, as shown in the 1970s~\citep{galtsov1972photoneutrino,skobelev1976reaction,DeRaad:1976kd,1970Ap&SS...7..407C,1970Ap&SS...9..453C}. We will follow a more recent treatment which corrects for several flaws of previous works~\citep{Kennett:1999jh}.

QED non-linear effects are of two kinds. On the one hand, a critical field $B_c=m_e^2/e=4.41\times 10^{13}\,\rm G$ is so large that the macroscopic field could create electron-positron pairs. We will always consider fields smaller than this value. The second class of effects is related to the vacuum birefringence 
 effect~\citep{Tsai:1975iz} as a large macroscopic field affects the refractive index, and can affect plasmon decay to neutrinos.\footnote{This is similar to the way the Cotton-Mouton effect is known to modify photon conversion to axions~\citep{Raffelt:1996wa}.} 
The magnetic field introduces an additional energy scale, the synchrotron frequency
 \begin{equation}
     \omega_B=\frac{e B}{m_e}
     = m_e \frac{B}{B_c}
     \simeq 11.5  \,B_{12}  \,\rm{keV}.
 \end{equation}
As a rule of thumb, one can compare how the magnetic field and the plasma modify the refractive index respectively. The equations for vacuum birefringence are given in~\cite{Tsai:1975iz}, and reduce for $B\lesssim B_c$ to $n_{\rm vacuum}^2-1\simeq \kappa\alpha^2 B^2/m_e^4$, where $\kappa$ is a constant which depend on the polarisation ($\kappa_{\parallel}=14/45$ and $\kappa_{\perp}=8/45$ for the electric field parallel and perpendicular to the external magnetic field respectively)~\citep{Raffelt:1996wa}.  As the refractive index in an unmagnetised plasma is $n^2_{\rm plasma}-1=-\omega_p^2/\omega^2$, one finds that vacuum birefringence is negligible for~\citep{Meszaros:1979xf,1980Ap&SS..73...33P}
\begin{equation}
    \frac{\alpha}{4\pi}\frac{\omega_B^2}{m_e^2}\frac{\omega^2}{\omega_p^2}
    = \frac{\alpha}{4\pi}\frac{B^2}{B_c^2}\frac{\omega^2}{\omega_p^2}
    \lesssim 1 ,
\end{equation}
which is safely satisfied for
$\omega \simeq T$, $B\lesssim B_c$ and the range of temperatures and densities considered here, see also Eq.~\eqref{PlasmaFrequenz}.

The most important effect that $B$ has in the range considered here is to force electrons on Landau levels with energies (notice that $\omega_B/m_e=B/B_c$)
\begin{eqnarray}
E_\nu = \sqrt{
p_\parallel^2
+
m_e^2\left(
1 + 2 \nu  \frac{B}{B_c}
\right)
}, 
\end{eqnarray}
with $p_\parallel$ the component of the momentum parallel to the magnetic field and
$\nu = l + \frac{1}{2} + \sigma$, $l\ge 0$ is an integer labelling the orbital angular momentum,
$\sigma=\pm\frac{1}{2}$, $g_0=1$, and all other $g_\nu=2$.
The electron density is then computed as (see~e.g.~\cite{1991ApJ...383..745L})
\begin{eqnarray}\label{newithLandaucylinders}
n_e = \frac{m_e^2}{(2\pi)^2}
\frac{B}{B_c}
\sum_{\nu=0}^{\infty}  g_\nu \int_{-\infty}^\infty d p_\parallel f^\nu_{-}(E_\nu), 
\end{eqnarray}
where $f_{\pm}^\nu = \left[\exp(\frac{E_\nu\pm \mu}{T}) + 1\right]^{-1}$ for respectively electrons ($-$) and positrons ($+$), and we neglected positrons.
In~\cite{Kennett:1999jh} the extreme case where only the lowest level is occupied was considered. We shall in the following investigate the impact of $B$ on the plasmon emission rate under this extreme assumption. By doing so we overestimate the effect of $B$ for realistic values in WDs, as one can see by estimating the magnitude of $B$ that would be needed to confine all electrons to the lowest Landau level. 
To show this we make the conservative assumption $T=0$, which yields a maximal (Fermi) momentum $p_\parallel=p_{F \parallel}(\nu)$ that is defined by the condition
\begin{eqnarray}\label{LandauLevelCondition}
 p_{F \parallel}^2(\nu) + m_e^2 \left(1 + 2\nu\frac{B}{B_c}\right) = \mu^2,
\end{eqnarray}
with $\mu$ the electron chemical potential. 
The highest occupied Landau level $\nu_{\rm max}$ corresponds to the maximal $\nu$ for which Eq.~\eqref{LandauLevelCondition} has a real solution in  $p_{F \parallel}$. Hence, if we require a value for $\nu_{\rm max}$, we can demand $p_{F \parallel}(\nu_{\rm max} + 1) = 0$ to identify the
 largest
  chemical potential $\mu$ for which only Landau levels up to $\nu_{\rm max}$ are filled, and express
\begin{eqnarray}
 n_e = \frac{m_e^3}{(2\pi)^2}
 \left(
 \frac{2B}{B_c}
 \right)^{3/2}\sum_{\nu=0}^{\nu_{\rm max}}
 g_\nu
 \sqrt{
1 + \nu_{\rm max} - \nu
 }.
\end{eqnarray}
Demanding $\nu_{\rm max}=0$ leads to
$n_e=m_e^3/(2\pi)^2 \ (2B/B_c)^{3/2}$
 or $B=\frac{1}{2}B_c(4\pi^2 n_e/m_e^3)^{2/3}$, i.e. $B/B_c\simeq 0.6$ for $Y_e\rho \simeq 10^6\, \rm g/cm^3$, where $Y_e$ is the number of electrons per baryon. 
Therefore, one needs incredibly large magnetic fields to keep the electrons in the lowest Landau level. We find that the value of the $B$ field for this to happen is much larger than the one cited in~\cite{Kennett:1999jh}, where it was assumed that the kinetic energy of the electron was due to the temperature, rather than the Fermi energy. However, for the strongest magnetic fields considered here, one still expects a sizeable fraction of the electrons in the lowest Landau levels.

As the general problem in which both the thermal motion of electrons (due to temperature or Fermi energy) and the magnetic fields are accounted for is very complicated~\citep{1986islp.book.....M,Kennett:1999jh}, we will limit ourselves to an electron cold plasma distribution, assuming that most of electrons are in the lowest Landau energy level. In this way, we will be able to estimate the maximal variation in the plasmon decay neutrino flux due to the presence of the magnetic field. In particular, the effect will be large on the axial-vector coupling emission: since this is associated to coherent spin oscillations (see also Appendix~\ref{App:plasmon}), the magnetic field breaks parity and aligns the spins, so that constructive interference is possible in the magnetised plasma for $\nu=0$.

Even with these simplifications, as the magnetic field breaks the isotropy of the medium, the problem of finding the dispersion relations of the electromagnetic field excitations becomes rapidly complicated~\citep{swanson2012plasma}. Therefore, we can estimate the effect the $B$ field has on neutrino emission from plasmon decay by considering the decay of plasmons propagating parallel or orthogonal to the magnetic field. The explicit evaluation of Eq.~\eqref{eq:Gamma-plasmon} is enormously simplified by using a long-wavelenght approximation for both the vector and axial-vector polarisation functions, and using the Stix form for the dielectric tensor~\citep{1962tpw..book.....S,1986islp.book.....M,swanson2012plasma} one can obtain simple expressions of the refractive index for different wave-vector directions and polarisations.

Taking these facts in combination, the different plasmon emission processes in a magnetised medium can be categorised according to $(i)$ whether they proceed via the axial-vector or vector coupling $(ii)$ the value of $\theta$ (i.e., the direction of propagation with respect to the magnetic field)
and $(iii)$ the polarisation of the mode in question which can be orientated relative to the direction of $B$. Expressions for general $\theta$ are quite complicated so in order to obtain some quantitative understanding of the size of effects, we examine various special cases below taken from~\cite{Kennett:1999jh}.

\paragraph{Circularly polarised parallel propagation ($\theta = 0$)}
  
    For $\theta=0$, the axial-vector and longitudinal contribution is not modified. The remaining contribution therefore proceeds via the vector coupling resulting in effect to a modification of $Q_T$. This consists of two circularly polarised modes: 
\begin{align}
\label{transv}
Q_{\pm}(\theta=0) =\frac{1}{4\pi} \left(\sum_{\nu_\alpha} C_{V}^{2}\right)\frac{G_F^2}{96\pi^4\alpha}\int_{\omega_{\rm min}}^\infty d\omega\ \omega^8 n_{\pm} (1 - n_{\pm}^2)^3 f_B(\omega),
\end{align}
where 
\begin{equation}
    n_{\pm }^2 =1-\frac{(\omega_p/\omega)^2}{1\pm \omega_B/\omega} ,
\end{equation}
and $\omega_{\rm min}$ corresponds to the refractive index $n=k/\omega$ being equal to 0, since below this cutoff the modes do not propagate.\footnote{Notice moreover that one should check that $n\le 1$, so that the excitations are timelike.} This expression coincides with the one found in~\cite{1970Ap&SS...7..407C}, and is independent of the axial-vector coupling.
Note that summing $Q_{\pm}$, taking the limit of zero magnetic field, and integrating over the plasmon direction, we find again again Eq.~\eqref{eq:qt}, as expected.

\paragraph{Perpendicular propagation ($\theta = \pi/2$), parallel polarisation}

A mode propagating perpendicular to $\textbf{B}$ but polarised parallel to $\textbf{B}$ is known as the ordinary mode (so-called because it has the same dispersion relation as though the anisotropy were not there). Both the axial-vector and vector contributions are modified by the $B$ field for this case, giving rise to 
\begin{align}
\label{ord1}
Q_o \left(\theta=\frac{\pi}{2}\right)
=\frac{1}{4\pi} \frac{G_F^2}{96\pi^4\alpha}\int_{\omega_{\rm min}}^{\infty} d\omega\  \omega^8 n_o (1-n_o^2)^2 \left[\left(\sum_{\nu_\alpha} C_{V}^{2}\right)(1 - n_o^2)+\left(\sum_{\nu_\alpha} C_{A}^{2}\right)n_o^2\right] f_B(\omega),
\end{align}
where the refractive index is independent of B,
\begin{equation}
n_{o}^2=1-\left(\frac{\omega_p}{\omega}\right)^2.
\end{equation}

\begin{figure}
\centering
    \includegraphics[scale=0.4]{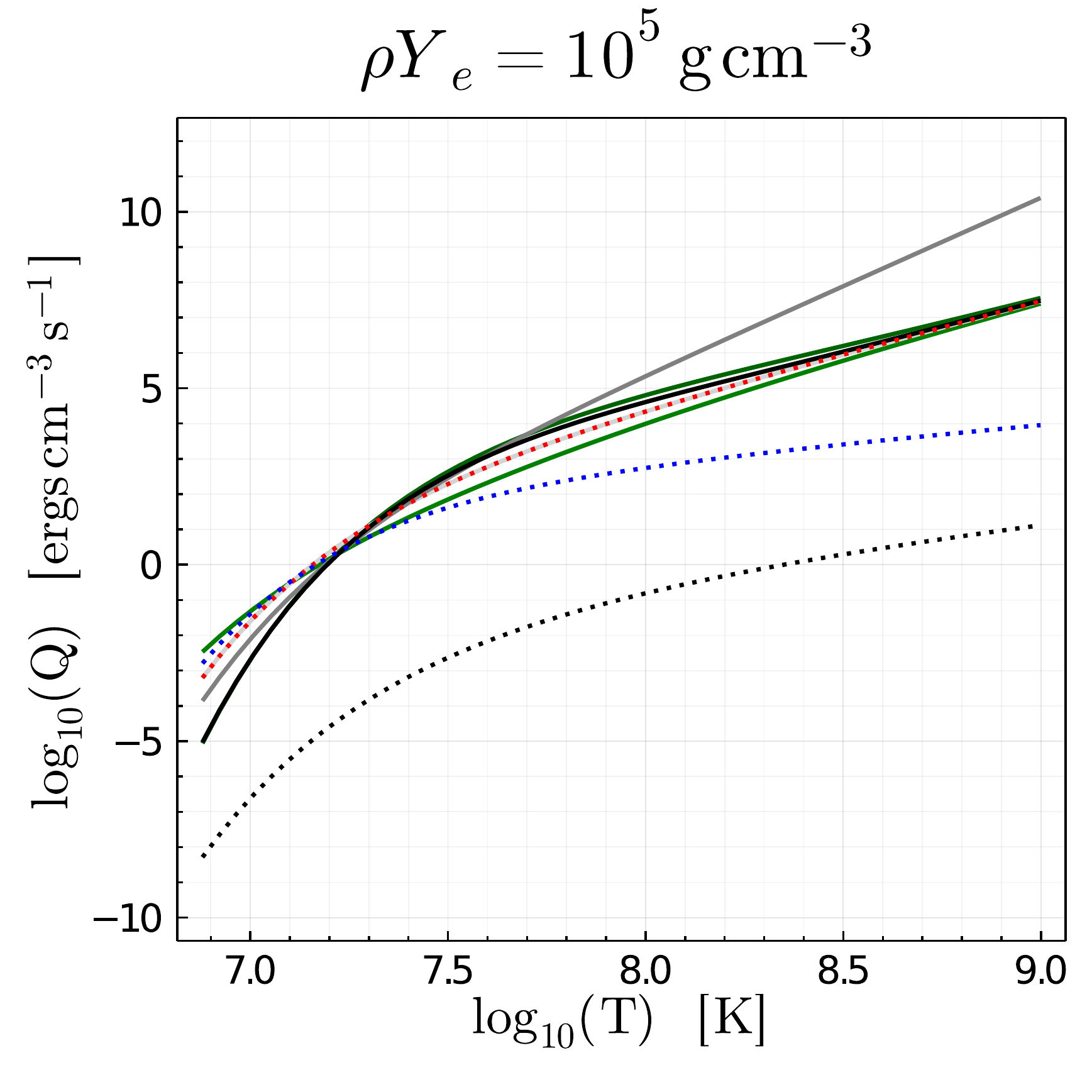}
     \includegraphics[scale=0.4]{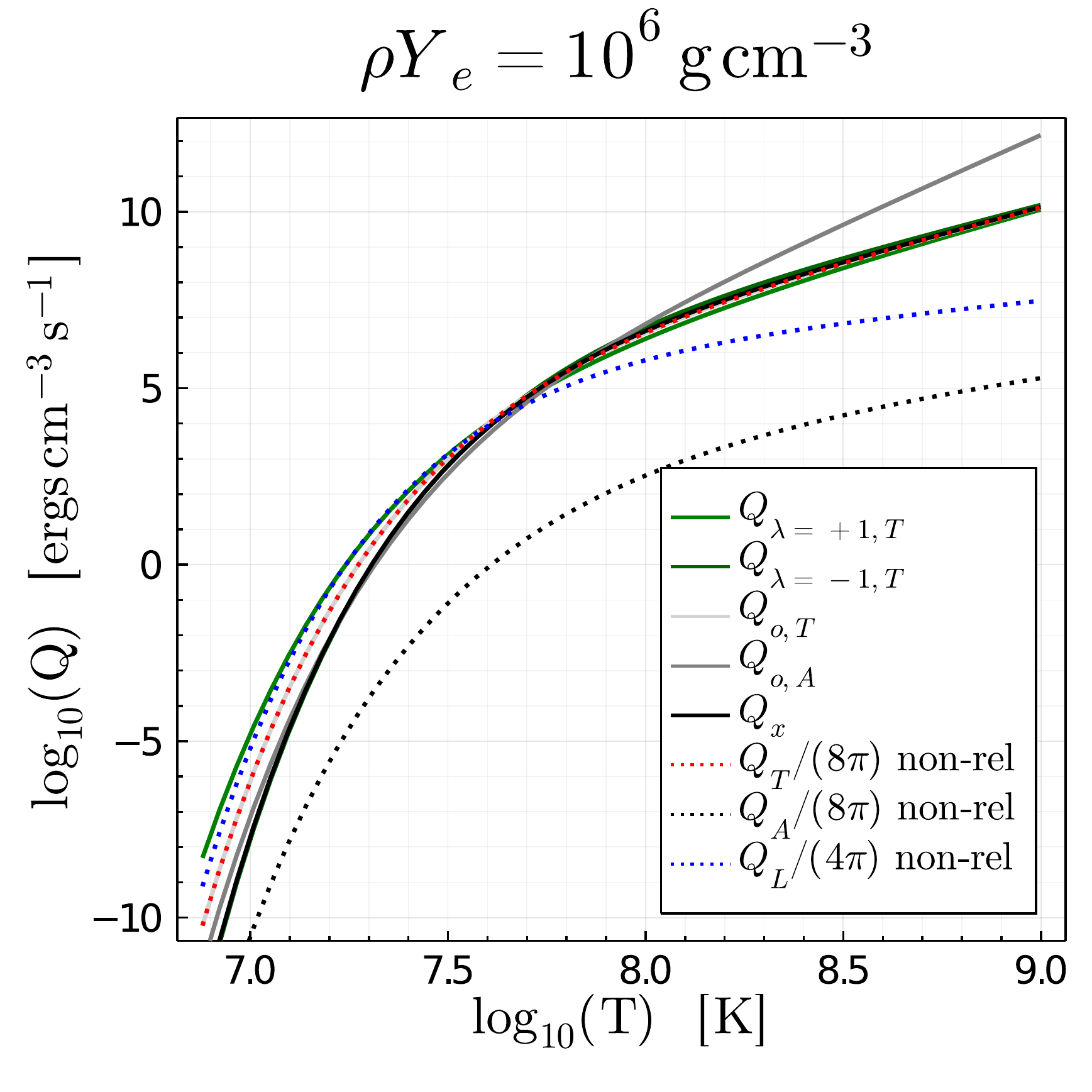}  \\
\includegraphics[scale=0.4]{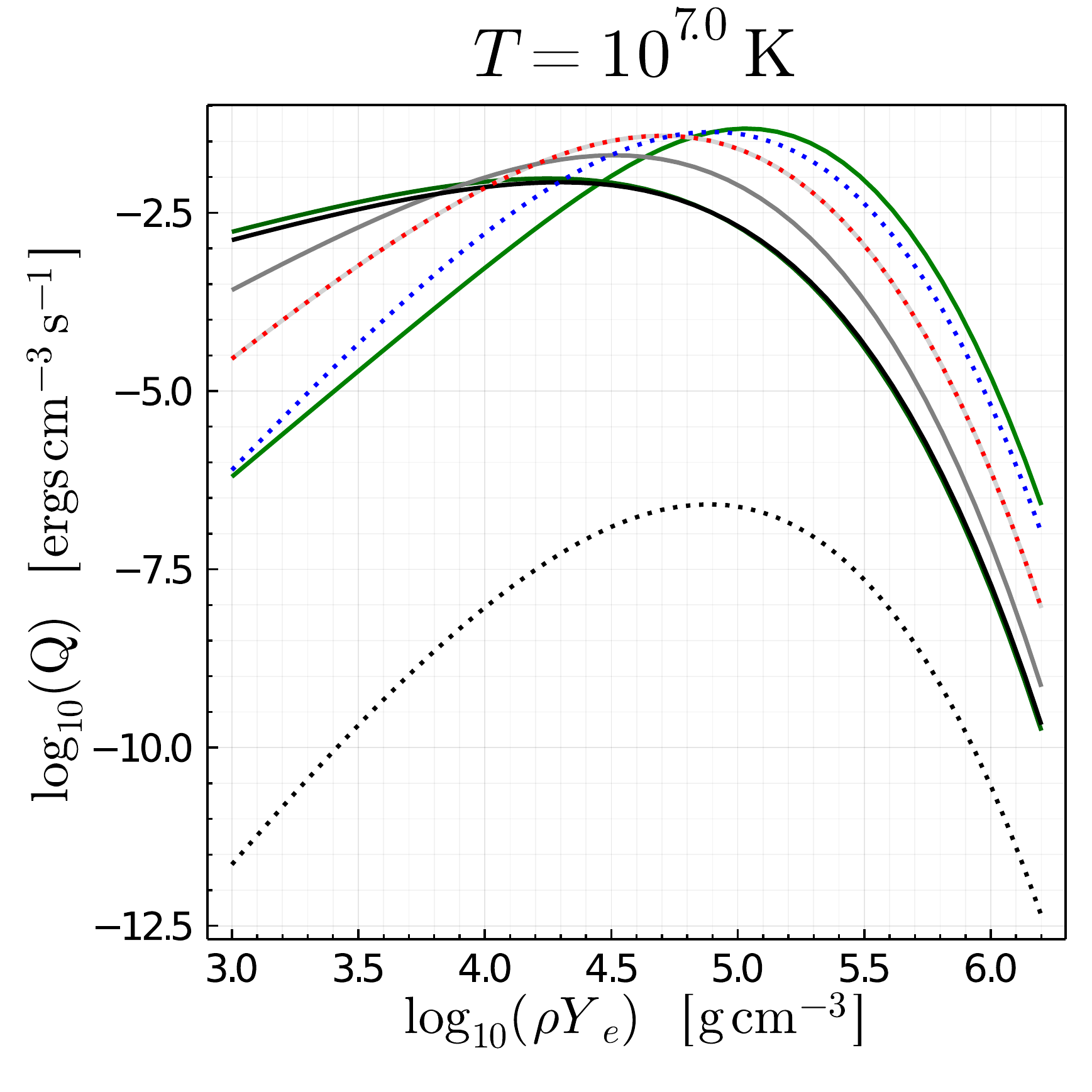}
    \includegraphics[scale=0.4]{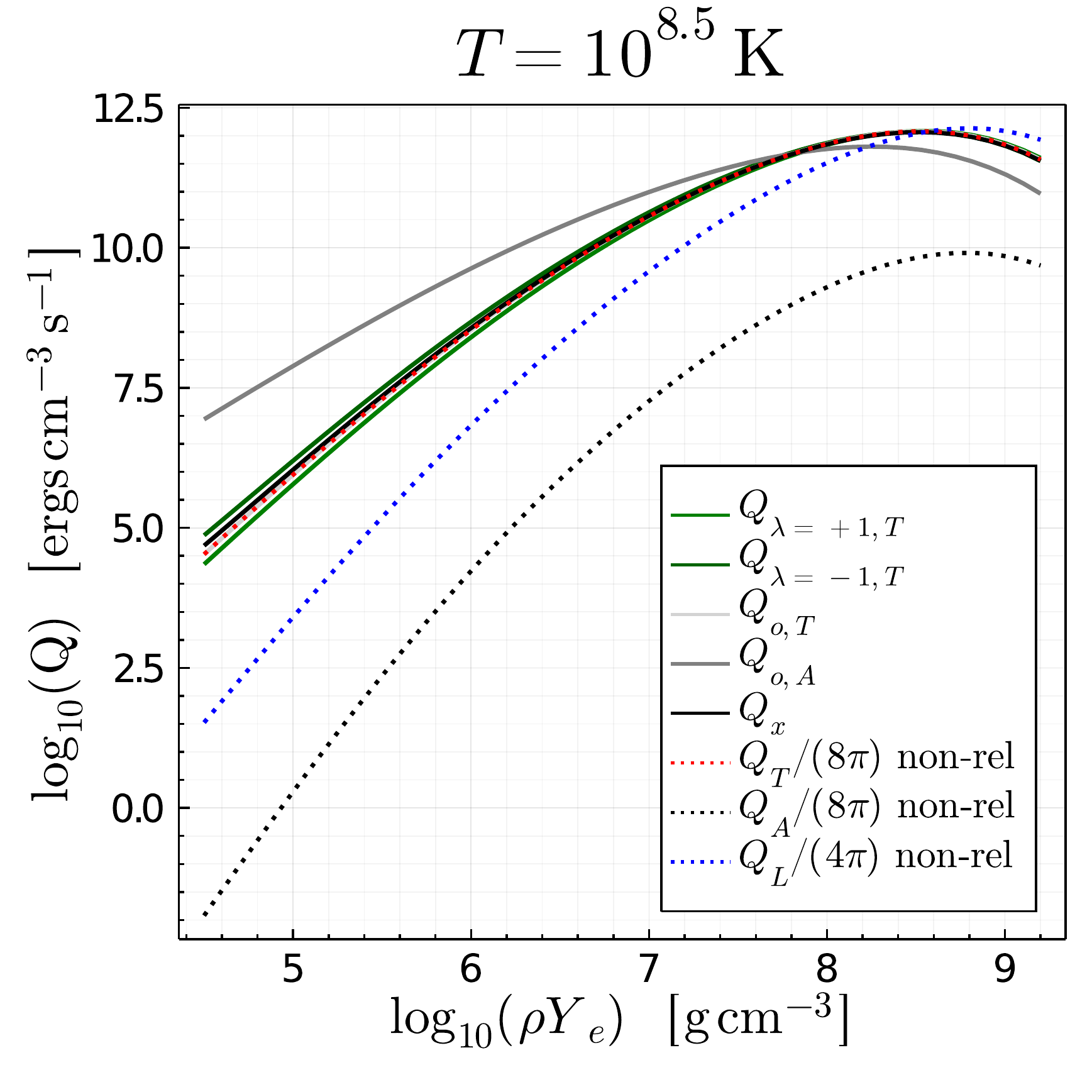}
    \caption{
    \textbf{Magnetised plasmon emission.} Effect of a strong magnetic field on the plasmon processes $\gamma \rightarrow \nu \bar{\nu}$. We show the energy loss per polarisation in a magnetised plasma ($Q_{\pm}, Q_{O},Q_{x}$) with solid lines, whilst the conventional plasmon processes in the unmagnetised medium ($Q_{T,A,L}$) are shown as dotted lines. We display the effect as a function of temperature (top row) for two densities and also as a function of varying density (bottom row) for fixed temperature and magnetic field $B = 7 \times 10^{11}$ G. 
    }
    \label{fig:axial}
\end{figure}

\paragraph{Perpendicular propagation ($\theta = \pi/2$), perpendicular polarisation}
This case is referred to as the extraordinary mode (the plasma dispersion relation receives magnetic field corrections), for which the vector (but not the axial-vector) contribution is modified:
\begin{align}
Q_x  \left(\theta=\frac{\pi}{2}\right)  =
\frac{1}{4\pi} \left(\sum_{\nu_\alpha} C_{V}^{2}\right)\frac{G_F^2}{96\pi^4\alpha}\int_{\omega_{\rm min}}^{\infty} d\omega\, \omega^8 n_x (1 - n_x^2)  \left[(S-1)^2 + D^2 - {\frac{4D^2 S(S-1)}{D^2 + S^2}}\right] f_B(\omega)
\end{align}
where
\begin{equation}
n_{x}^2=\frac{S^2-D^2}{S} ,\qquad S=1-\frac{(\omega_p/\omega)^2}{1 - (\omega_B/\omega)^2},
\qquad
D=\frac{\omega_B\omega_p^2}{\omega(\omega^2-\omega_B^2)} .
\end{equation}
We summarise the size of these contributions in Fig.~\ref{fig:axial}, which compares the plasmon processes in a magnetised medium to the plasmon processes in the unmagnetised medium. 
Neutrinos from plasmon decay might potentially constrain the internal $B$ field of WDs. The colder part of the WDLF shows some excessive cooling~\citep{Giannotti:2015kwo}, while the hot part does not show it~\citep{Hansen:2015lqa}. Notice however that the WDLF uncertainty is large in the hottest range. From Fig.~\ref{fig:axial} we see that the plasmon decay is mostly modified at rather large temperatures.
We will show in the next section that the emission from synchrotron radiation dominates plasmon decay for values of the magnetic field whose effect on plasmon decay is large.
If in the future this part of the WDLF will be measured precisely, a dedicated and computationally more challenging estimate of the $B$ field effect on plasmon decay would be demanded featuring, e.g., a warm plasma~\citep{Kennett:1999jh}.

\subsection{Neutrino pair synchrotron radiation}\label{Sec:NSR}
\begin{figure}
\plottwo{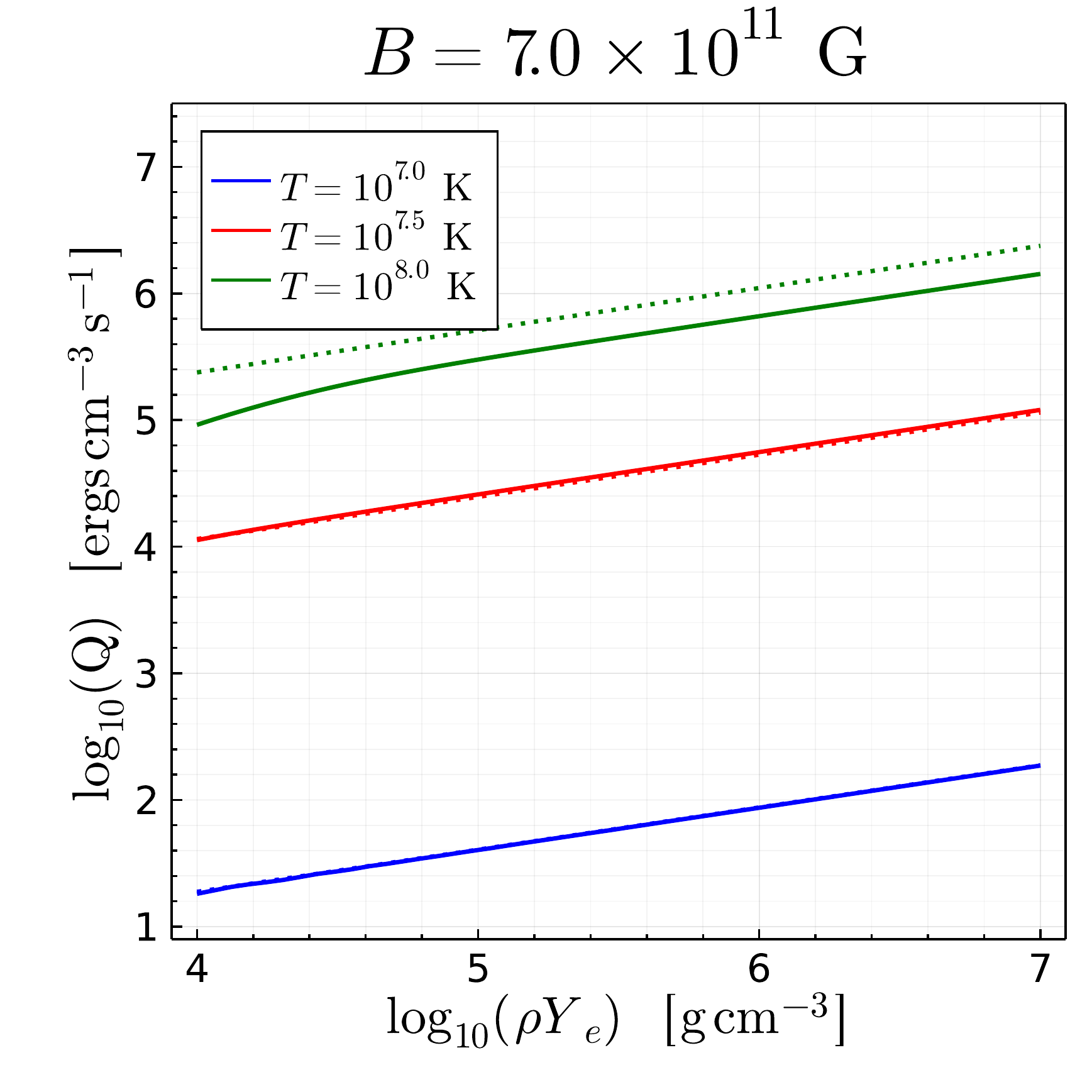}{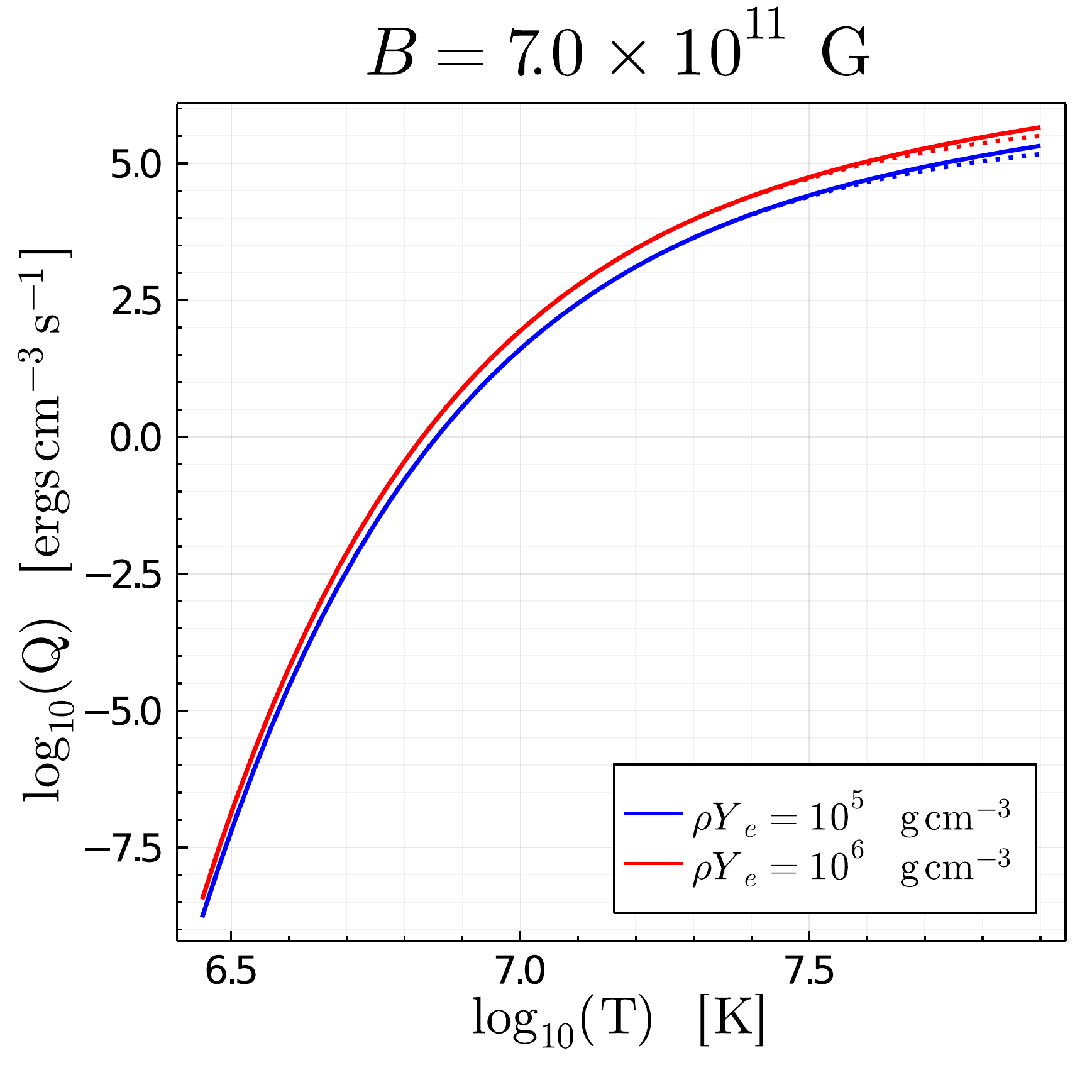}
\plottwo{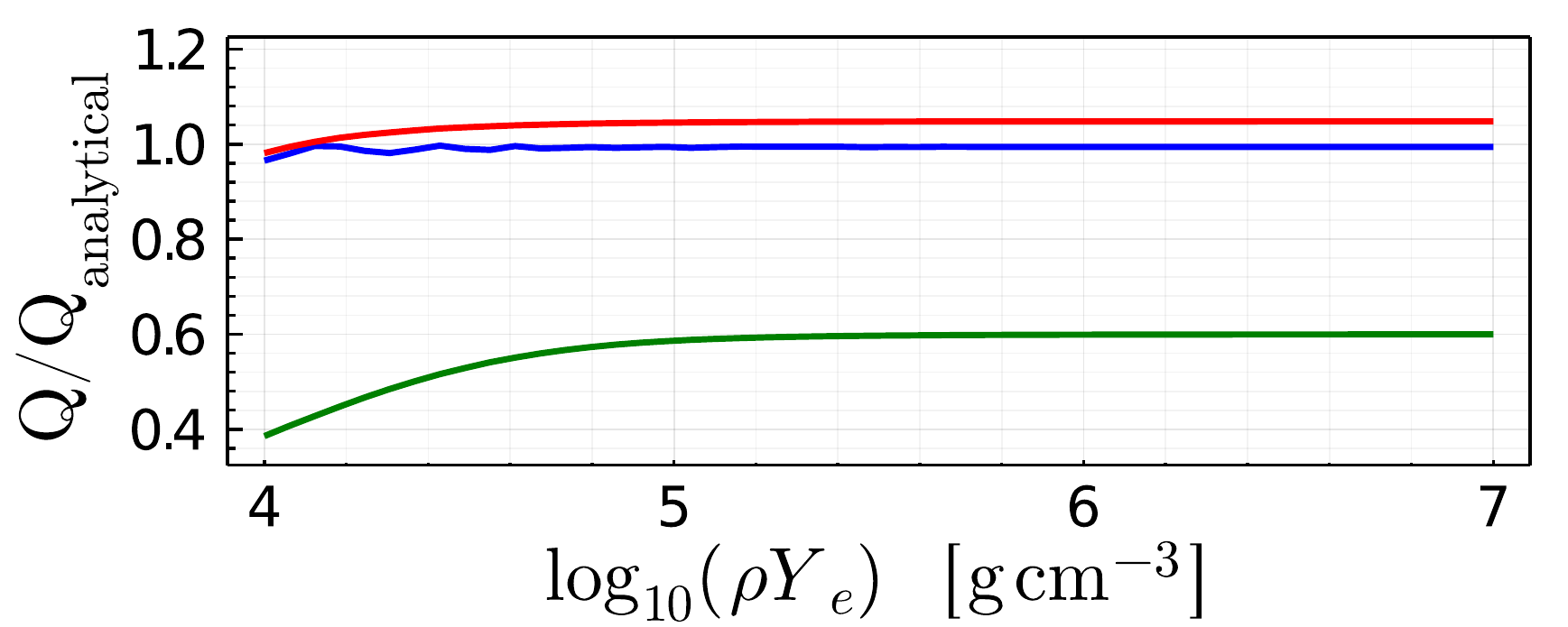}{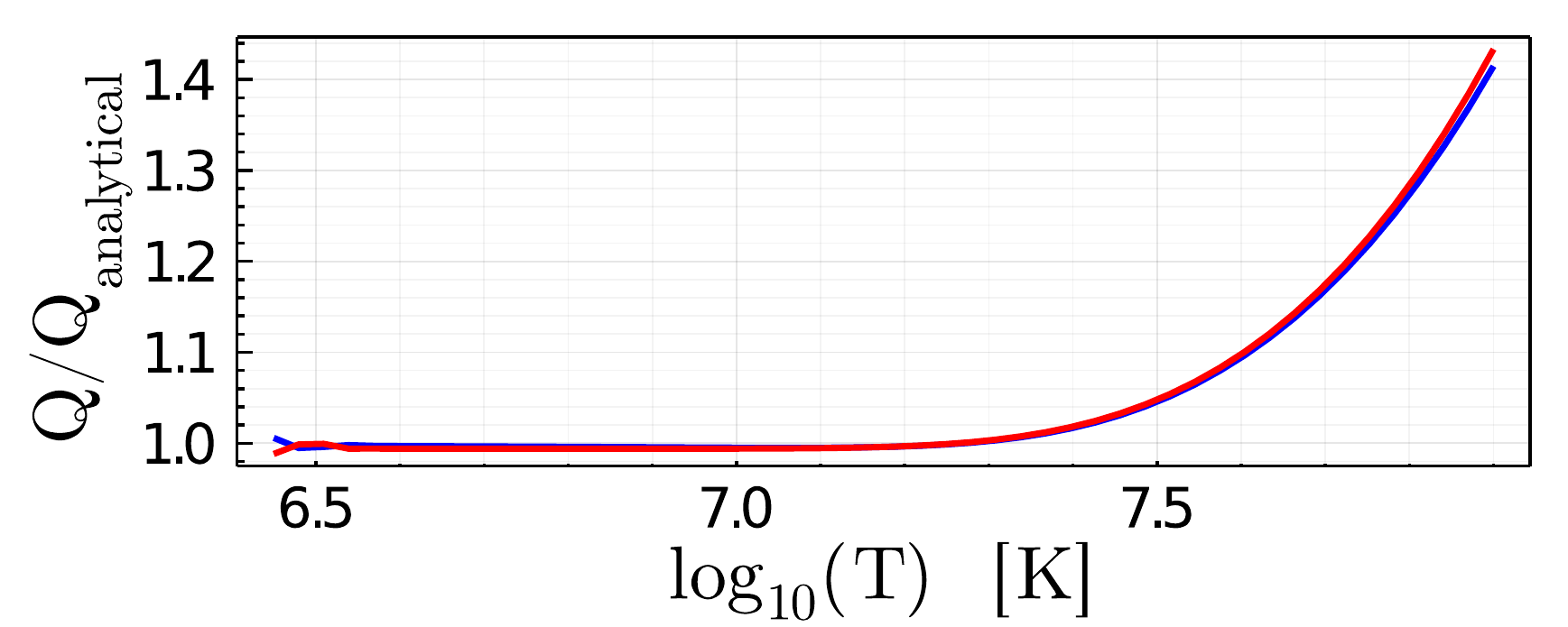}
	    \caption{\textbf{Synchrotron Emission.} Full numerical result (solid lines) Eq.~\eqref{eq:numsync} for the synchrotron emissivity as a function of density (left panel) and temperature (right panel). For comparison we also display the analytic estimates (dotted lines) discussed in the text. In the left panel we show the estimates for weak quantisation Eq.~\eqref{QSynchWeakQuant} (red and blue dotted) and non-quantisation Eq.~\eqref{QSynchNonQquant} (green dotted). In the right panel we show Eq.~\eqref{QSynchWeakQuant} and Eq.~\eqref{QSynchWeakQuant} at low and high temperatures, respectively.
	    }
	    \label{fig:Synchrotron1}
\end{figure}

\begin{figure}
	    \centering
	    \epsscale{0.5}
	\plotone{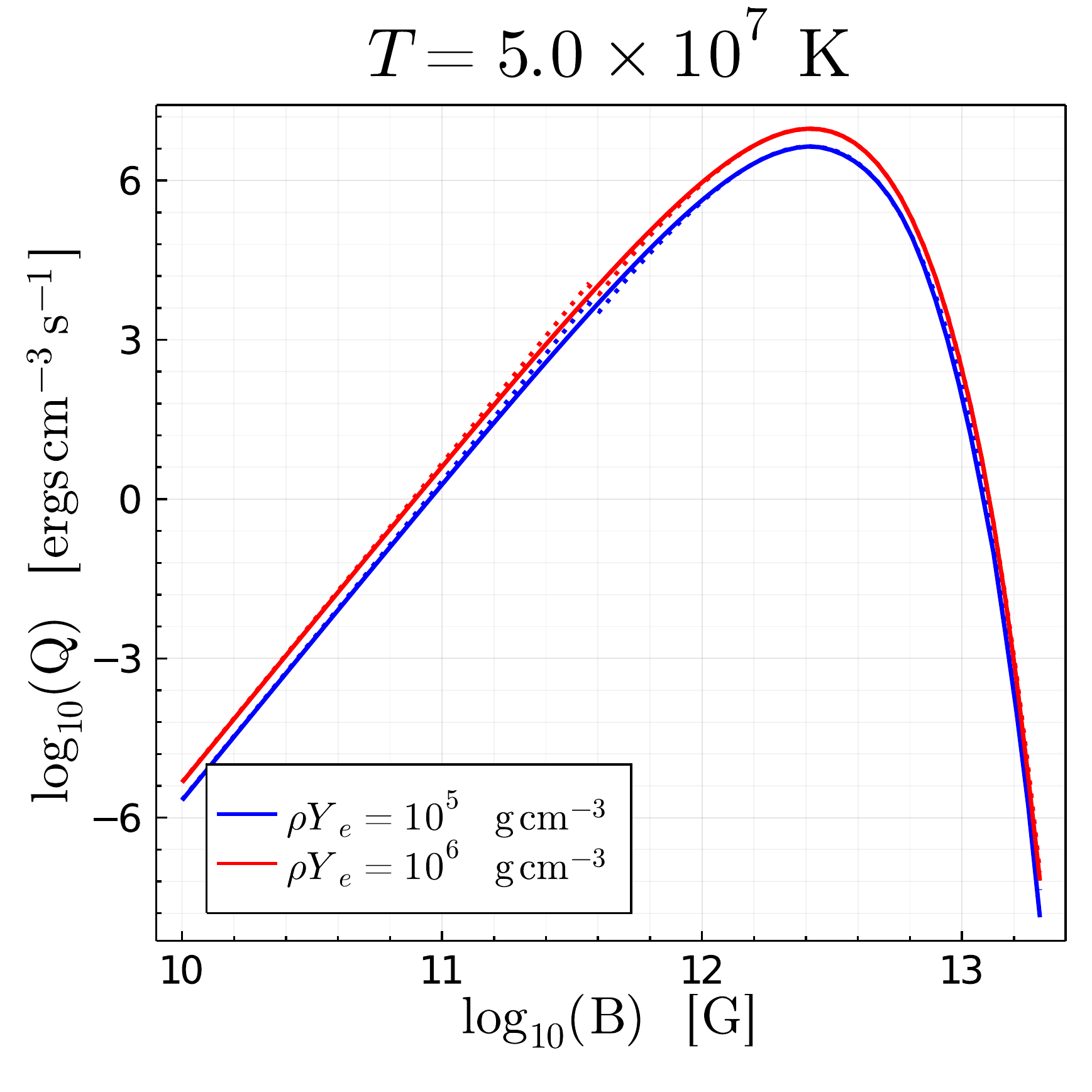}
	    \caption{\textbf{Magnetic dependence of synchrotron emission.} Full numerical result Eq.~\eqref{eq:numsync} for synchrotron emission as function of $B$ for various densities. As before we show the analytic limits  Eq.~\eqref{QSynchWeakQuant} and Eq.~\eqref{QSynchNonQquant}.  }
	    \label{fig:Bdep_Synch}
\end{figure}

Similarly to neutrino production from bremsstrahlung, which is the main neutrino production channel during the surface cooling phase, synchrotron emission is possible thanks to an external field. The $B$ field forces the electrons to rotate around the magnetic field lines. The electron momentum is not conserved, allowing for the process $e\xrightarrow{{B}}e \, \bar{\nu} \, \nu$. 
Neutrino pair synchrotron emission has been the subject of many studies, spanning several decades (see e.g.~\cite{Landstreet:1967zz,1981AN....302..167I,Yakovlev:2000jp}). Using the notation introduced in Section~\ref{Sec:plasmonmagnetic}, the emissivity reads \citep{Kaminker:1992su}
\begin{equation}\label{eq:numsync}
		Q_\text{syn} = \left(\sum_{\nu_\alpha} C_{A}^{2}\right) \frac{G_F^2 \omega_B^6}{30 \pi^3 } \frac{m\omega_B}{\left(2\pi \right)^2} \sum_{\nu=1}^\infty \int_{-\infty}^{+\infty} dp_{\parallel} \left[ f_{-}^\nu \left(1 - f_{-}^{\nu-1}\right) +  f_{+}^\nu \left(1 - f_{+}^{\nu-1}\right)\right],
	\end{equation}
	where as above
	\begin{equation}
		f_{\pm}^\nu = \left[\exp(\frac{E_\nu\pm \mu}{T}) + 1\right]^{-1}, \qquad \qquad 
		E_{\nu} = \sqrt{p_{\parallel}^2 + m_e^2 + 2\nu\omega_B m_e}.
	\end{equation}
Notice that all neutrino flavors are equally produced by the synchrotron process. It is useful to obtain two formulae which describe synchrotron emission in regime relevant for WDs. We assume $T\ll m_e$, so that positrons can again be neglected ($f_+^{\nu}\ll f_-^{\nu}$). The degenerate limit $T \ll p_F^2/(2m_e)= E_F^{\mathrm{NR}}$ is valid, as $p_F \simeq (3 \pi^2 n_e)^{1/3}$ for $\omega_B \ll E_F \simeq \mu$, and it is always large in WD cores (see Appendix~\ref{App:plasmon}). The emissivity is
\begin{subequations}
\begin{align}
 \quad Q_\text{syn}(\omega_B \ll T \ll E_F^{\mathrm{NR}}) &= \left(\sum_{\nu_\alpha} C_{A}^{2}\right) \frac{G_F^2 \omega_B^6}{60 \pi^3 }   \frac{3}{2} \frac{T}{E_F^{\rm NR}} n_e ,\label{QSynchNonQquant} \\
Q_\text{syn}( T\ll \omega_B  \ll E_F^{\mathrm{NR}})  & =  \left(\sum_{\nu_\alpha} C_{A}^{2}\right) \frac{G_F^2 \omega_B^6}{60 \pi^3 } \frac{3}{2} \frac{\omega_B}{E_F^{\rm NR}}e^{- \omega_B/T} n_e . \label{QSynchWeakQuant}
\end{align}
\end{subequations}
 These limits are referred to as \textit{non-quantised degenerate} and \textit{weakly quantised degenerate}, respectively. In Figs.~\ref{fig:Synchrotron1} and \ref{fig:Bdep_Synch} we display the full numerical results Eq.~\eqref{eq:numsync} together with the analytic estimates Eq.~\eqref{QSynchNonQquant} and Eq.~\eqref{QSynchWeakQuant}. The analytical approximation is precise up to large temperatures, where the energy loss should be interpolated between Eq.~\eqref{QSynchNonQquant} and Eq.~\eqref{QSynchWeakQuant}. Given the strong dependence on the magnetic field, the analytical approximation of Eq.~\eqref{QSynchWeakQuant} is good enough to describe the region of transition from non-quantised to quantised, considering our accuracy goals. Synchrotron emission is compared against all other processes in Fig.~\ref{fig:all_processes}, from which we see that synchrotron emission can enhance the energy loss from WDs.

 	\subsection{Heating from magnetic field decay}\label{Sec:Ohmic}
 	
Magnetic fields as large as those discussed in the previous sections can have additional effects on the evolution of a WD. 
 	One of them is the heating of the star due to Ohmic decay of the magnetic field~\citep{Cumming:2002vq,Ferrario2015}.
 	The resulting conversion of the magnetic field energy\footnote{Note that we define the electromagnetic field energy density $U_{\rm em}$ using the particle physics friendly rationalised units rather than the Gaussian \textit{unrationalised} units of~\cite{chandrasekhar1953problems}, i.e.
\begin{equation}
U_{\rm em}=\frac{B^2}{2},
\end{equation}
rather than $U_{\rm em}=\frac{B^2}{8\pi}$ for the energy density of the $B$-field. One can thus
compute everything using natural units, and eventually substitute $1.95\times 10^{-2} \, \rm eV^2\rightarrow 1 \, \rm G$, see also the Appendix A of~\cite{Raffelt:1996wa}.
} 
$U_{\rm em} =  B^2/2$ into thermal energy is the dominant process in the evolution of a field smaller than $10^{12}\,\rm G$~\citep{Heyl:1998cz}. 
 We do not attempt at quantifying the potential effect of Ohmic heating on the surface temperature at later stages~\citep{valyavin2014suppression}, but instead only estimate it at the level of energy gain and loss per unit volume in the core in order to compare it to other processes discussed in this work.
 	 As an example we consider a hypothetical magnetic field in the well-known WD G117-B15A,
 	for which we take the parameters~\citep{Kepler:2000pd,BischoffKim:2007ve}
 	\begin{equation}\label{G117B15Aparameters}
 	Y_e\rho\simeq 10^6 \ {\rm g/cm^3}, \ R= 9.6\times10^8\ {\rm cm}, \ T=1.2 \times 10^7\ {\rm K}. 
 	\end{equation}
 	Assuming an and exponential decay $B=B_0\exp(-t/t_{\rm Ohm})$
 	with $t_{\rm Ohm}\simeq 3\times 10^{11}$ years (\cite{Cumming:2002vq,Ferrario2015}) and an age much smaller than $t_{\rm Ohm}$, the energy density deposited by an initial magnetic field $B_0\simeq 3\times 10^{11}\,\rm G$ per unit time is
 	$-Q_{\rm Ohm} = dU_{\rm em}/dt 
 	\simeq B_0^2/t_{\rm Ohm}\simeq 750\, \rm erg/cm^3/s$.
Integrated over the volume, this corresponds to $\simeq 7 \times 10^{-4} L_\odot$.
Comparing this to the energy loss due to neutrino cooling obtained from Eq.~\eqref{QSynchWeakQuant} with Eq.~\eqref{G117B15Aparameters}, 
one finds $|Q_{\rm Ohm}/Q_\text{syn}|\simeq 7.6$, and hence, the additional heating by Ohmic decay of the $B$ would clearly dominate over the enhancement of the cooling rate caused by the $B$ field. 
Moreover, the temperature would be larger due to the heating produced in previous times as well.

\section{\textbf{Applications and Discussion}}	

\begin{figure}
    \plottwo{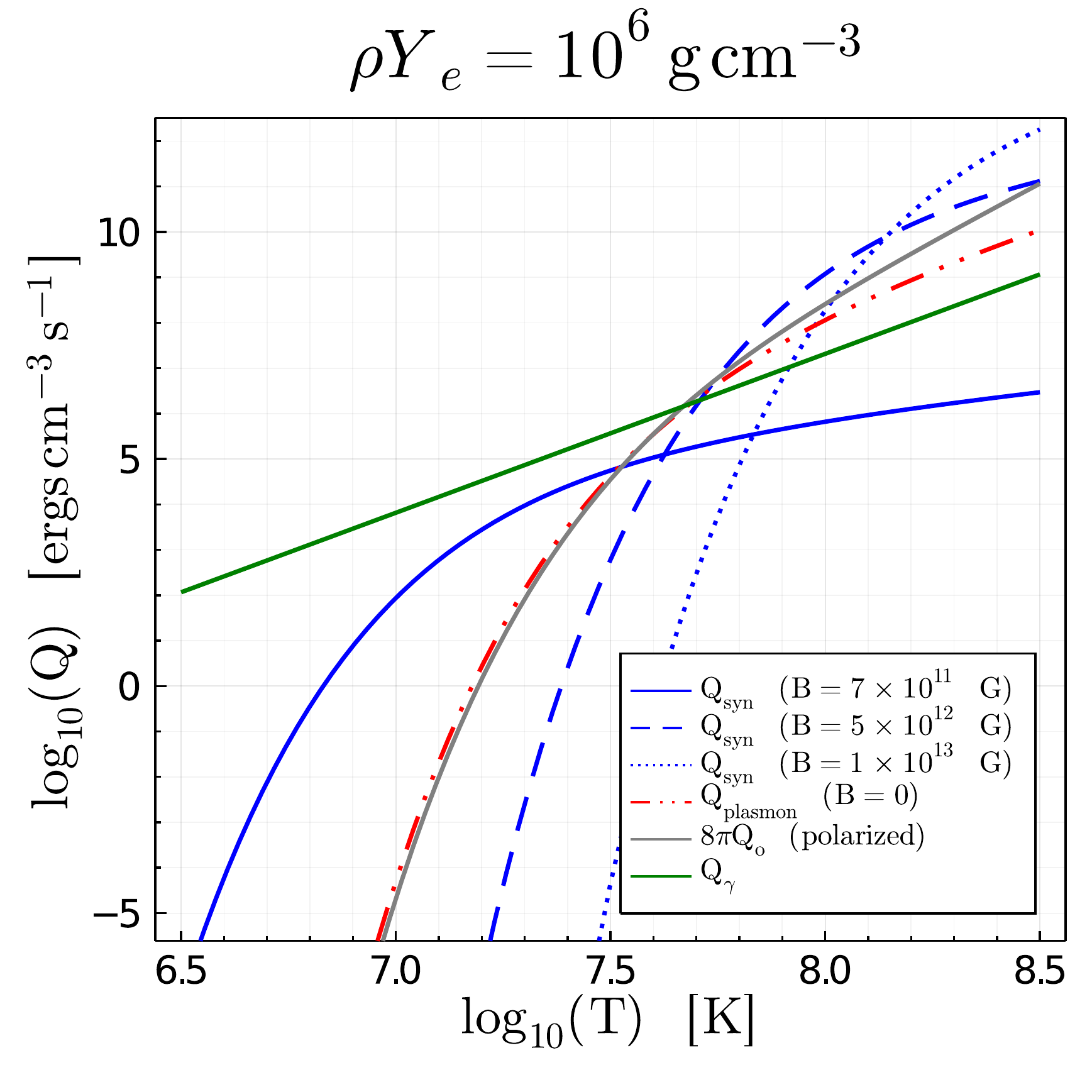}{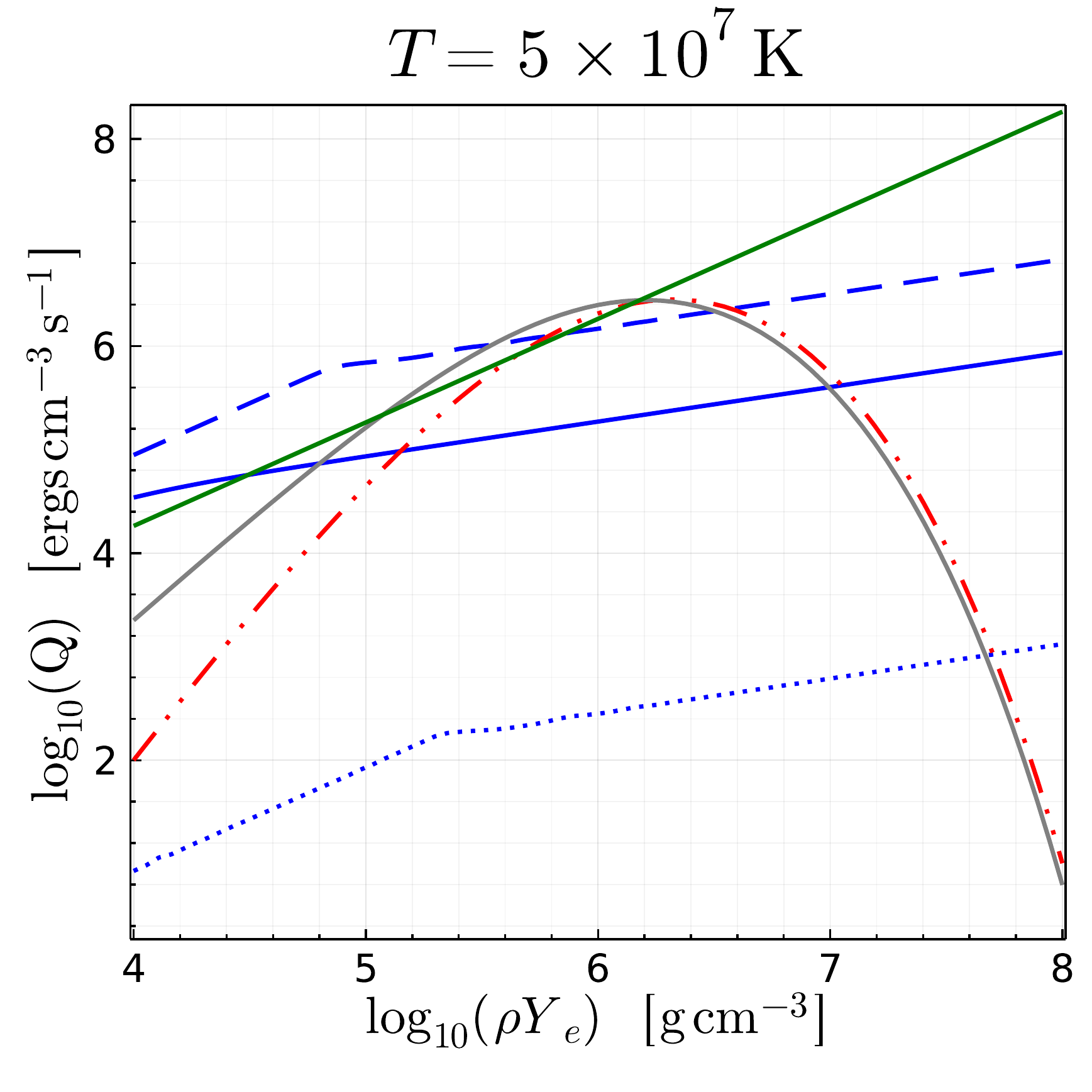} 
    \caption{\textbf{Magnetic fields and WD cooling.}  Magnetic field induced cooling processes, including the emissivities due to synchrotron emission and to plasmon decay in a polarised medium. For comparison we also show the plasmon decay emissivity in the unmagnetised plasma, and the photon surface emissivity. Note that photon emission actually depends on the global properties of the star so $Q_\gamma$ here is a volume average.  We show the behaviour as a function of temperature (left panel) and of density (right panel).
    } 
    \label{fig:all_processes}
\end{figure}

 	 We show in Fig.~\ref{fig:all_processes} all the cooling processes we discussed in the paper. Notice that at low temperatures neutrino bremsstrahlung is the dominant processes in neutrino production~\citep{Winget:2003xf}. However, this is a moot point, since at these temperatures, photon surface emission (which we display) becomes the dominant cooling process. 
 	We also notice in passing that the axial-vector contribution to plasmon decay in presence of a magnetic field $B\simeq 5 \times 10^{12}\, \rm G$ is subdominant compared to neutrino pair synchrotron radiation.

As a first application, we can notice that $n_e=Y_e\rho/m_u$ where $m_u$ is the atomic mass unit. Therefore, the energy-loss rate 
due to synchrotron emission
per mass unit reads, using the weakly quantised limit,
 	\begin{equation}
 		\epsilon_\text{syn}^\text{WD} \simeq \left(\sum_{\nu_\alpha} C_{A}^{2}\right) \frac{G_F^2 \omega_B^6 Y_e}{60 \pi^3 m_u }\frac{3}{2}\frac{\omega_B}{E_F^{\mathrm{NR}}}\exp[-\omega_B/T] .
 	\end{equation}
	Assuming $Y_e=0.5$, $\rho=2 \times 10^5 \, \rm g/cm^3$, and $T\simeq 10^8 \, \rm K\simeq 8.6 \, \rm keV$, typical parameters for a RG core, we need to ask the energy-loss rate per mass unit to be smaller than $10 \,\rm erg\, g^{-1}s^{-1}$, to prevent the cooling from delaying the onset of helium burning in the core (see Chapter~2 of~\cite{Raffelt:1996wa}). Intriguingly, recent works have advanced the hypothesis of an ubiquitous magnetic field in RG cores~\citep{2016ApJ...824...14C}. We find that neutrino cooling constrains such a ubiquitous field to be smaller than the (arguably gargantuan) value $B_{\rm max}\lesssim 10^{12}\,\rm G$, where the approximation of negligible effect on the global structure might be already broken.

 	As a second application, let us consider the secular variation of their period of pulsation of stars like G117-B15A, which have been suggested to show an excessive cooling, prompting an axion production interpretation~\citep{Isern:1992gia,Giannotti:2015kwo}.\footnote{It has been argued that these hints  point to an axion solution also because they are produced with a certain temperature dependence~\citep{Giannotti:2015kwo}.} 
Using the parameters of Eq.~\eqref{G117B15Aparameters}, 
 	we find that a field $B_{\rm max}\simeq 3\times 10^{11}\,\rm G$ produces a cooling of about $10^{-4}L_\odot$, which roughly matches the excessive cooling observed~\citep{Giannotti:2015kwo}. We stress that this interpretation of the cooling hint is far from conclusive, as the magnetic field should be totally confined in the core, something difficult to achieve~\citep{Peterson:2021teb}.
 	A more careful analysis is demanded, but it could be important to further investigate a SM solution to the cooling hints, particulary in light of recent works, based on X-ray signals from axions reconverting in the magnetosphere of MWDs, challenging the axion hypothesis for this cooling hint~\citep{Dessert:2021bkv}.

 	As a third application we investigate the potential to use the impact of $B$-fields on neutrino emission to impose an upper bound $B_{\rm max}$ on the internal magentic field in the WD population. 
 	Let us assume that all WDs have a magnetic field buried underneath their surface. For simplicity, we describe all the WDs as spheres with a fixed density and electron to baryon ratio $ \rho =2\times 10^6 \,\rm g/cm^3$ and $Y_e=0.5$. Moreover, we assume the magnetic field to have the same magnitude over the entire stellar core, with a sharp drop near the surface. If $B_{\rm max}\simeq 1.2 \times 10^{12}\,\rm G$, the neutrino energy loss, obtained from Eq.~\eqref{QSynchWeakQuant}, is larger than that of surface emission, given by Eq.~\eqref{VolumeAveragedQgamma}, for temperatures larger than $3.3\times 10^7 \rm \, K$. A value $B_{\rm max}\simeq 6\times 10^{11}\,\rm G$ in WDs implies an energy loss $10\%$ of the surface emission at a core temperature of $2\times 10^7 \, \rm K$. This corresponds very roughly to a $10\%$ effect on the WDLF.
 	The argument above can be made slightly more precise by integrating the emissivities over a WD model, rather than assuming a constant value for the density profile (see Appendix~\ref{App:profile}).
 	
The effect of synchrotron radiation on stellar evolution could be analyzed more precisely through some code such as MESA~\citep{Paxton:2010ji}, together with experimentally determined WDLF. This would give more realistic constraints, properly including the density profile evolution of the WDs, as well as their mass distribution. The numerical implementation is possible thanks to Eqs.~\eqref{QSynchNonQquant} and \eqref{QSynchWeakQuant}, which can be directly fed to a stellar evolution code, as the error due to the use of analytical approximations is never larger than a factor $\mathcal{O}(1)$ at intermediate temperatures. Therefore, the analytical approximations are precise enough, since synchrotron radiation depends on a large (sixth) power of $B$. We leave this analysis for a future study.

Finally, as a very simple estimate, we compare the Ohmic heating
 $Q_{\rm Ohm}$ studied in section \ref{Sec:Ohmic} 
 to  $Q_\gamma$ in \eqref{VolumeAveragedQgamma} and demand that the heating due to Ohmic decay is less than $10\%$ of the energy loss due to surface cooling (a number that would be of comparable magnitude as the cooling anomaly quoted in \cite{Giannotti:2015kwo}, but of opposite sign). This translates into a bound $B < 2.73 \times 10^{11} {\rm G}/\sqrt{Y_e}$ for the parameters in Eq.~\eqref{G117B15Aparameters}.
 	Notice however that heating depends on the Ohmic decay time, and the bound on $B$ roughly scales as $\propto \sqrt{t_{\rm Ohm}}$. 
 	Therefore, if the lifetime of the magnetic field is larger than expected, the neutrino cooling bound applies, while if $t_{\rm Ohm}\simeq 3\times 10^{11}$~years is a good estimate, the magnetic field is primarily constrained by the anomalous heating it would produce in this case. Unless a finely tuned cancellation happens between the heating due to the magnetic field decay and the cooling due to neutrino synchrotron radiation, these two constraints are complementary.


\section{Conclusions}\label{Sec:conclusions}

Besides affecting the global quantities of degenerate stars, strong magnetic fields could modify their cooling.  
A large magnetic field can catalyze the cooling by modifying the ordinary processes of a non-magnetised core~\citep{DeRaad:1976kd,skobelev1976reaction,galtsov1972photoneutrino,1970Ap&SS...7..407C,Kennett:1999jh,1970Ap&SS...9..453C}, and by introducing additional processes as the neutrino pair synchrotron radiation~\citep{Landstreet:1967zz,1981AN....302..167I,Kaminker:1992su}. 
We  studied the impact of magnetic fields in the interior of RGs and WDs on neutrino cooling rates, which are constrained both at the population level via the WDLF and in individual stars for some WDs by pulsation measurements, and derived a new limit on the magnitude of the interior $B$-fields.
This independent limit 
complements several recent observations. Astrometric measurements suggest that most of RGs could host very large magnetic fields~\citep{Fuller_2015,Stello_2016,2016ApJ...824...14C}. Similarly WDs could also hide large magnetic fields beneath their surface~\citep{2016ApJ...824...14C}.

The problem of quantifying the impact that magnetic fields in stellar cores have on 
observable quantities factorises into two parts:
firstly identifying the relevant \emph{microphysical} processes and computing their contributions to the emissivity, and secondly investigating the consequences that this has on the \emph{macroscopic} structure of the WD and its evolution.
In the present work we focussed on the first aspect.
We found that the strongest effect on the cooling is due to synchrotron radiation, rather than to the modification of plasmon decay.
Even without a detailed analysis of the stellar evolution this leads to a stronger constraint on the $B$ field than the well-known stability requirement. 
While this improvement only amounts to a factor of order one, the two constraints are complementary, as they depend on different observables, showing once more how particle production in stars can be a diagnostic tool for astrophysics.

More specifically, we found that if an ubiquitous magnetic field is present in RGs, it cannot be larger than $10^{12}\,\rm G$. From a comparison to the WDLF in the simple form of the Mestel's cooling law, we have shown that an ubiquitous field hiding beneath WD surfaces needs to be smaller than $10^{12}\,\rm G$ and could be potentially limited down to $6\times 10^{11}\,\rm G$, slightly improving bounds coming from a stability requirement. Moreover, we found that variable stars like G117-B15A, which show excessive cooling,  could be potentially affected by neutrino pair synchrotron radiation. However, we stress that the interpretation of the cooling hints in terms of an internal magnetic field can be challenged, as such large magnetic fields should potentially show up in asteroseismological surveys.

In addition to the impact that $B$-fields can have on WD cooling, their Ohmic decay can also heat the star's core. 
The energy gain due to this process can potentially exceed the loss from cooling by synchrotron radiation.
Hence, the non-observation of excess heating can also impose an upper bound on $B$, which we estimate to be a factor 2-3 stronger than the bound from the non-observation of excessive cooling due to synchrotron emission  (assuming no cancellation between the two effects, which in reality would of course always co-exist). A quantitative study, however, crucially relies on the details of $B$-dissipation due to Ohmic damping, which goes beyond the scope of this work. The different parametric dependencies of the energy gain and loss caused by large $B$-fields make these bounds complementary.

Our work could be expanded in several directions. One could study numerically the effect of synchrotron radiation on stellar evolution (through some code such as MESA~\citep{Paxton:2010ji}) with experimentally determined WDLF to obtain a more precise bound on the internal magnetic field. 
Constraining the population of highly magnetised WDs would not only help to better understand stellar evolution, but e.g.~also help to estimate the gravitational wave background expected from very massive WDs \citep{Kalita:2019cio}.
Finally, the production of some
new elementary
particles can be enhanced in the presence of large magnetic fields, thus providing an additional signature~\citep{Caputo:2020quz,OHare:2020wum,Caputo:2021kcv}.
For example, axion-like particles  with an electron coupling can be produced via synchrotron radiation~\citep{Kachelriess:1997kn}. Such particles, if discovered, would be open new window on the internal magnetic fields of WDs.

\section*{Acknowledgments}
EV thanks Georg Raffelt, Javier Redondo, Aldo Serenelli, and Irene Tamborra on discussions about related topics, and Matteo Cantiello for an important discussion on magnetic fields in stellar cores.
We thank Matteo Cantiello, Georg Raffelt, and Irene Tamborra for comments on a first version of the paper.
Finally, we thank the anonymous referee for bringing the Ohmic heating to our attention.
JIM is  supported  by  the  F.R.S.-FNRS  under the  Excellence  of  Science  (EOS)  project  No.  30820817(be.h). The work of EV was supported was supported in part by the U.S. Department of Energy (DOE) Grant No. DE-SC0009937.

\appendix
\section{Plasmon decay}\label{App:plasmon}
We recollect here the main results of~\cite{Braaten:1993jw} who found expressions for plasmon decay in isotropic and homogeneous plasmas for any condition. 
 
\subsection{Full relativistic results}

The electron and positron distributions are
	\begin{equation}
		f_-(E)=\frac{1}{e^{(E-\mu) / T}+1}, \quad f_+(E)=\frac{1}{e^{(E+\mu) / T}+1},
	\end{equation}
	where $E=\sqrt{p^{2}+m_{e}^{2}}$, so that the net charge density
	\begin{equation}
		n_c =\frac{1}{\pi^{2}} \int_{0}^{\infty} d p\, p^{2}\left(f_-(E)-f_+(E)\right)
	\end{equation}
depends only on the chemical potential $\mu$ and the temperature $T$, which are therefore the two parameters on which the QED plasma depends on when ion features are negligible. The dispersion relations for plasmons depend on the typical electron velocity
\begin{equation}
		v_{*}=\frac{\omega_{1}}{\omega_{p}},\quad v = \frac{p}{E},
	\end{equation}
	where the plasma frequency is given by
		\begin{equation}
		\omega_{p}^{2}=\frac{4 \alpha}{\pi} \int_{0}^{\infty} d p \frac{p^{2}}{E}\left(1-\frac{1}{3} v^{2}\right)\left(f_-(E)+f_+(E)\right),
	\end{equation}
	and
	\begin{equation}
		\omega_{1}^{2}=\frac{4 \alpha}{\pi} \int_{0}^{\infty} d p \frac{p^{2}}{E}\left(\frac{5}{3} v^{2}-v^{4}\right)\left(f_-(E)+f_+(E)\right) .
	\end{equation}

The dispersion relations for transverse and longitudinal plasmons with energy $\omega$ and momentum $k$ are respectively
	\begin{equation}
		\omega_{t}^{2}=k^{2}+\omega_{p}^{2} \frac{3 \omega_{t}^{2}}{2 v_{*}^{2} k^{2}}\left(1-\frac{\omega_{t}^{2}-v_{*}^{2} k^{2}}{\omega_{t}^{2}} \frac{\omega_{t}}{2 v_{*} k} \log \frac{\omega_{t}+v_{*} k}{\omega_{t}-v_{*} k}\right), \quad 0 \leq k<\infty
	\end{equation}
and
	\begin{equation}
		\omega_{l}^{2}=\omega_{p}^{2} \frac{3 \omega_{l}^{2}}{v_{*}^{2} k^{2}}\left(\frac{\omega_{l}}{2 v_{*} k} \log \frac{\omega_{l}+v_{*} k}{\omega_{l}-v_{*} k}-1\right), \quad 0 \leq k<k_{\max }
	\end{equation}
	where $k_{\max }$ is the momentum at which the longitudinal plasmon dispersion relation crosses the light cone,
	\begin{equation}
		k_{\max }=\left[\frac{3}{v_{*}^{2}}\left(\frac{1}{2 v_{*}} \log \frac{1+v_{*}}{1-v_{*}}-1\right)\right]^{1 / 2} \omega_{p} .
	\end{equation}

The total energy loss due to plasmon decay is given by
	\begin{equation}
		Q_{\gamma\rightarrow\nu\bar\nu} = Q_T + Q_A + Q_L
	\end{equation}
	where the contribution from transverse plasmons through vector and axial-vector couplings, and from longitudinal plasmons through vector coupling, are respectively
	\begin{subequations}
	\begin{align}\label{eq:QTFUllApp}
		Q_T &= 2\left(\sum_{\nu_\alpha} C_{V}^{2}\right) \frac{G_F^2}{96\pi^4 \alpha} \int_0^{+\infty} dk\, k^2 Z_t(k) \left(\omega_t^2 - k^2\right)^3 f_B(\omega_t)
\\
		Q_A &= 2\left(\sum_{\nu_\alpha} C_{A}^{2}\right) \frac{G_F^2}{96\pi^4 \alpha} \int_0^{+\infty} dk\, k^2 Z_t(k) \left(\omega_t^2 - k^2\right) \Pi_A(\omega_t, k)^2 f_B(\omega_t)
\label{eq:QLFUllApp}
\\
		Q_L &=\left(\sum_{\nu_\alpha} C_{V}^{2}\right) \frac{G_F^2}{96\pi^4 \alpha} \int_0^{k_\text{max}} dk\, k^2 Z_l(k) \omega_l^2 \left(\omega_l^2 - k^2\right)^2 f_B(\omega_l)
\end{align}
\end{subequations}
	where the plasmon distribution is given by
	\begin{equation}
		f_B(\omega) = \frac{1}{e^{\omega/T}-1} .
	\end{equation}
	To compute them, one needs the functions renormalizing the coupling~\citep{Braaten:1993jw,Raffelt:1996wa},
	\begin{equation}
		Z_{t}(k)=\frac{2 \omega_{t}^{2}\left(\omega_{t}^{2}-v_{*}^{2} k^{2}\right)}{3 \omega_{p}^{2} \omega_{t}^{2}+\left(\omega_{t}^{2}+k^{2}\right)\left(\omega_{t}^{2}-v_{*}^{2} k^{2}\right)-2 \omega_{t}^{2}\left(\omega_{t}^{2}-k^{2}\right)}
	\end{equation}
	for the transverse plasmon, and
	\begin{equation}
		Z_{l}(k)=\frac{2\left(\omega_{l}^{2}-v_{*}^{2} k^{2}\right)}{3 \omega_{p}^{2}-\left(\omega_{l}^{2}-v_{*}^{2} k^{2}\right)}
	\end{equation}
for the longitudinal one.
The axial-vector polarisation function reads
			\begin{eqnarray}
			 \Pi_A = \frac{2 \alpha}{\pi}\frac{\omega^2 - k^2}{k} \int_0^\infty dp \frac{p^2}{E^2}
			  \left(
			  \frac{\omega}{2 v k} \log \frac{\omega + vk}{\omega - vk} - \frac{\omega^2 - k^2}{\omega^2 - v^2 k^2}
			  \right)
			  \left(f_-(E)-f_+(E)\right),
			\end{eqnarray}
where
	\begin{equation}
		\omega_{A}=\frac{2 \alpha}{\pi} \int_{0}^{\infty} d p \frac{p^{2}}{E^{2}}\left(1-\frac{2}{3} v^{2}\right)\left(f_-(E)-f_+(E)\right) .
	\end{equation}

	\subsection{Non-relativistic limit of plasmon cooling}\label{nonrelaplasmon}

In the non-relativistic limit, the dispersion relations for transverse and longitudinal plasmons are respectively\footnote{Notice that transverse plasmons have (in this limit) a dispersion relation resembling ones of a massive particle.  On the other hand, the dispersion relation of the longitudinal plasmon is such that its energy is independent of its momentum. Therefore, even in the non-relativistic limit the comparison of plasmons to massive particles should not be taken too literally~\citep{Vitagliano:2017odj}.
}
\begin{equation}
\omega^2_t=
    \omega_p^2+\bk^2
\qquad\hbox{and}\qquad
  \omega^2_l=\omega_p^2 \, 
\end{equation}
The plasma frequency in the limit $T\ll m_e$ is given in terms of the electron density
$n_e$ by
\begin{equation}\label{PlasmaFrequenz}
    \omega_p^2=\frac{4\pi\alpha\,n_e}{m_e}\left[1+\frac{1}{m_e^2}(3\pi^2 n_e)^{2/3}\right]^{-1/2}
    \,
    \simeq \left(20\,\rm keV \rho_6^{1/2}\right)^2
    \end{equation}
where $\rho_6=\rho/10^6 \,\mathrm{g}\,\mathrm{cm}^{-3}$, and we assumed $Y_e=0.5$.  We shall also make use of the approximations
(valid for $T \ll m_e$)  $Z_t \simeq1$, $\Pi_A \simeq \omega_{p}^2/2m_e k( \omega^2 - k^2)/\omega^2 $. 

With this in mind we obtain the following approximations: 
\begin{align}
		Q_T^{\rm NR} &\simeq 2\left(\sum_{\nu_\alpha} C_{V}^{2}\right) \frac{G_F^2 \omega_p^6}{96\pi^4 \alpha} \int_0^{+\infty} dk\, k^2  f_B(\omega_t)\simeq\left(\sum_{\nu_\alpha} C_{V}^{2}\right) \frac{G_F^2 \omega_p^6}{24\pi^4 \alpha}\zeta(3)T^3, \label{eq:qt}
		\end{align}
				where in the second steps we approximate $\omega_t\simeq k$, and
\begin{align}
	 Q_L^{\rm NR} & \simeq \left(\sum_{\nu_\alpha} C_{V}^{2}\right) \frac{G_F^2}{96 \pi^4 \alpha} \int_0^{\omega_{ p}} dk k^2 \omega_{p}^2 (\omega_{ p}^2 - k^2)^2 f_B(\omega_{ p})
	 \simeq \left( \sum_{\nu_\alpha} C_{V}^2\right)
	 \frac{G_F^2}{96 \pi^4 \alpha} \, \frac{T}{\omega_{ p}} \, \frac{8 \omega_{p}^9}{105},
	\end{align}
where in the second step we assume $\omega_p\ll T$.	Notice however that for our conditions $\omega_p\simeq T$, so one should use the first equality as a good approximation.
	The axial-vector contribution is
	\begin{align}\label{eq:qaNR}
		Q_A^{\rm NR} &= 2\left(\sum_{\nu_\alpha} C_{A}^{2}\right) \frac{G_F^2\omega_p^2}{96\pi^4 \alpha} \int_0^{+\infty} dk\, k^2  \Pi_A(\omega_t, k)^2 f_B(\omega_t)\nonumber\\
	&=2\left(\sum_{\nu_\alpha} C_{A}^{2}\right) \frac{G_F^2\omega_p^6}{96\pi^4 \alpha} \int_0^{+\infty} dk\, \frac{k^4}{\omega^4_t} \frac{\omega_p^4}{4m_e^2} f_B(\omega_t).
	\end{align}
We show the approximations Eqs.~\eqref{eq:qt}--\eqref{eq:qaNR} in Fig.~\ref{fig:plasmon} from which we conclude that relativistic effects can be reasonably neglected for the regimes relevant for WDs. 

 That the agreement with non-relativistic results is good is unsurprising, as can be seen by making the following estimates. If the WD is to be non-relativistic and degenerate then $p_F \simeq (3 \pi^2 n_e)^{1/3}$, which gives
\begin{equation}\label{eq:pf}
    p_F\simeq 515 \, \mathrm{keV} (Y_e \rho_{_6})^{1/3},
\end{equation}
with $\rho_{_6}$ expressed in units $\rm 10^6 \, g/cm^3$ and $Y_e$ is the number of electrons per baryon. The latter are related by $n_e = Y_e n_c = Y_e \rho/m_u$ where $\rho$ is the mass density and $m_u$ is the atomic mass unit. This corresponds  (taking $Y_e \rho = 10^6 \, \rm g/cm^3$) to a non-relativistic Fermi energy  $E_F^{\mathrm{NR}} = p_F^2/(2m_e) \simeq 200 \, {\rm keV}$  (and it is actually even smaller, as the average density of the profile is about $15\%$ of the central density~\citep{Shapiro:1983du}). 
The highest temperatures of relevance here are $T \simeq 10^8\,\rm K=8.621\, \rm keV$. Hence we have that $E_F^{\mathrm{NR}} \ll m_e$ and $T \ll m_e $.

These approximations also allow us to understand the relative sub-dominance of the axial contribution. If we neglect logarithmic corrections, we see that $Q_T/Q_A=\mathcal{O}({m_e^2 T^2/\omega_p^4})$, therefore finding that in a non-relativistic and non-magnetised plasma the emission through the axial-vector coupling is highly suppressed. As discussed in~\cite{Raffelt:1996wa} (see also~\cite{Vitagliano:2017odj}), one can heuristically think of the plasmon decay as dipole radiation from electrons coherently oscillating in space, thus contributing to vector emission, while the axial-vector emission corresponds to electron spin flips. 
As spins do not oscillate coherently in an unpolarised medium, the emission rate is suppressed. This also implies that in the non-magnetised medium plasmon decay produces mostly $\nu_e\bar\nu_e$ pairs~\citep{Vitagliano:2017odj}.

\section{Virial theorem, stability and magnetic field strength}\label{App:stability}

The virial theorem provides a general recipe to relate the time averaged (denoted $\braket{\cdot}$) total kinetic energy of a system of particles, held together by interactions, with that of the total potential energy of the system,
\begin{equation}
    2\langle T_{\rm tot}\rangle=n\langle V_{\rm tot}\rangle 
\end{equation}
where $T$ is the kinetic energy and $V\propto r^{n}$ is the interaction bounding the system. For gravity $V=\Omega\propto r^{-1}$, and one obtains a well known result in mechanics. The generalisation to a stellar environment which includes a magnetic field is immediate. One finds~\citep{chandrasekhar1953problems,Shapiro:1983du,Coelho_2014}
\begin{equation}
    2\langle T\rangle+3(\gamma-1)\langle U\rangle+\langle \Mag\rangle+\langle \Omega\rangle=0 .
\end{equation}
Here, we separated the macroscopic motion of the fluid, $T$, from the kinetic energy associated to the temperature, $3(\gamma-1)\langle U\rangle$, where $\gamma$ is defined by the polytrope of tha gas, $P=K \rho^\gamma$, and $\rho$ is the density of the system. As we can see, the generalisation simply requires us to substitute $\Omega$ with $\Omega+\Mag$, where $\Mag$ is the magnetic field energy.
If there is no macroscopic motion and $\langle T\rangle=0$, the virial theorem requires
\begin{equation}
  3(\gamma-1)\langle U\rangle+\langle \Mag\rangle+\langle \Omega\rangle=0 ,
\end{equation}
which combined with the definition of total energy
\begin{equation}
 \langle E\rangle= \langle U\rangle+\langle \Mag\rangle+\langle \Omega\rangle,
\end{equation}
gives the requirement for dynamical stability
\begin{equation}
(3\gamma-4)(|\langle \Omega\rangle|-\langle \Mag\rangle )>0 .
\end{equation}
In absence of a magnetic field, a configuration is stable as far as $\gamma>4/3$. The presence of the magnetic field spoils stability once the energy associated to the magnetic field is larger than the gravitational potential. Explicitly we have
 \begin{equation}
\Omega=-\frac{3}{5-n}\frac{G_{N} M^2}{R}, \qquad 
\Mag=  \frac{1}{2} \int d^3 \textbf{x}\,  \braket{B^2(\textbf{x},t)}  
\end{equation}
where we defined $n=1/(\gamma-1)$. Let us now find the maximum magnetic field which can be supported within some sub-region of the star whilst maintaining stability.

Typically, magnetic field strengths can vary significantly between the outer layers and core of a WD (see e.g. \cite{Kalita:2019cio}). As stated in the main text, determining the precise magnetic field structure would entail modelling beyond the scope of this work. Nonetheless the following toy setup is highly effective in illustrating how small magnetised core regions can affect cooling whilst remaining compatible with stability. Let us consider some prototype magnetic field profile
\begin{equation}
B = 
\left\{
\begin{array}{ll}
B_{\rm core}   & \quad 0 \leq r \leq  R_{\rm core}\\
B_{\rm outer}   & \quad  R_{\rm core} \leq r \leq R 
\end{array}
\right. 
\end{equation}
with $B_{\rm core} \gg B_{\rm outer}$.  Then, an upper bound on the strength of the core magnetic field comes from virial balance ($|\braket{\Omega}| = \braket{\Mag}$) would give (in the non-relativistic limit $\gamma=5/3$),
\begin{equation}\label{DeathMagnetic}
B_{\rm core} \leq 
\frac{M}{R^2} \sqrt{\frac{ 18 G_N}{4 \pi (5-n)}}
\left(\frac{R}{R_{\rm core}}\right)^{3/2}
 = 2.4 \times 10^8 \frac{M}{M_\odot}\left(\frac{R_\odot}{R}\right)^2 \, \left(\frac{R}{R_{\rm core}}\right)^{3/2}\, \mathrm{G} 
\end{equation}
where we assumed $M_\odot = 1.99\times10^{33}\, \mathrm{g}$ and $R_\odot=6.96 \times 10^{10}\, \mathrm{cm}$. We assume $R_{\rm core}\simeq R$ in the following, with the magnetic field hidden just below the surface.

Let us now consider a canonical WD with a mass of around $0.6 M_\odot$. From an observational point of view, this is well justified, as the WDs have a mass distribution peaked at a low mass~\citep{1984A&A...132..195W,2007MNRAS.375.1315K} due to the combined effect of the progenitor initial mass function favoring small masses~\citep{2001tass.book.....P}, and to the non-linear progenitor-WD mass relation, which also tends to favor WDs with small masses~\citep{2013MmSAI..84...58K}. One also avoids the problems associated with the change in structure of the large WD, due to the relativistic degenerate gas of electrons having $\gamma\simeq 4/3$, which makes it very sensitive even to small magnetic fields~\citep{Shapiro:1983du}.

 For non-relativistic electrons, the mass-radius relation is given by the simple result 
(see Eq.(2.1) of~\cite{Raffelt:1996wa}, or Section~3.3 of~\cite{Shapiro:1983du})
\begin{equation}
R=10,500 \, \mathrm{km}\,  (0.6 M_\odot/M)^{1/3}(2 Y_e)^{5/3},
\end{equation}
and the central density found to be
\begin{equation}
\rho_{\rm core}=1.46 \times 10^{6}\,\mathrm{g}\, \mathrm{cm}^{-3}(M/0.6 M_\odot)^{2}(2 Y_e)^{-5} ,
\end{equation}
where the mean molecular weight of the electrons is approximately $Y_e\simeq 0.5$. Therefore, we find that the maximum $B$ field allowed in WDs is
\begin{equation}\label{eq:B_On_SHell}
B_{\rm max}= 8.8 \times 10^{11} (M/0.6 M_\odot)^{1/3}\,\rm G . 
\end{equation}
In principle, the magnetic field in the core could be larger than the one shown in Eq.~\eqref{eq:B_On_SHell}. The density profile is a monotonically decreasing function of the radius, so a top-hat configuration could allow a larger magnetic field when applying the virial theorem to the core alone. Moreover, an analytical estimate is unavoidably
 rough, and does not account for several effects, which can potentially hinder the stability requirement bound. For example, Hamada and Salpeter~\citep{1961ApJ...134..683H} showed as early as 1961 that Coulomb corrections and chemical composition can affect the mass-radius relations at the few tens percent level. Most importantly, a more consistent treatment of the electron equation of state suggests that fields as large as $10^{13}\, \rm G$ for a $0.7 M_\odot$ do not destabilize the star, and actually have a negligible effect on the mass-radius relation~\citep{Suh:1999tg}. Other authors have found smaller sustainable fields~\citep{Franzon:2015gda,Bera:2014wja,Chatterjee:2016szk}.  Therefore, we conclude that the bound on the internal magnetic field of WDs due to stability requirement is somewhere between $B_{\rm max}\lesssim 10^{12}\, \rm G$ and $B_{\rm max}\lesssim 10^{13}\, \rm G$ with variablity coming (amongst other things) from the spatial compactness of the higher fields within the star.  

\section{Integration over the white dwarf profile}\label{App:profile}

\begin{figure}
    \plottwo{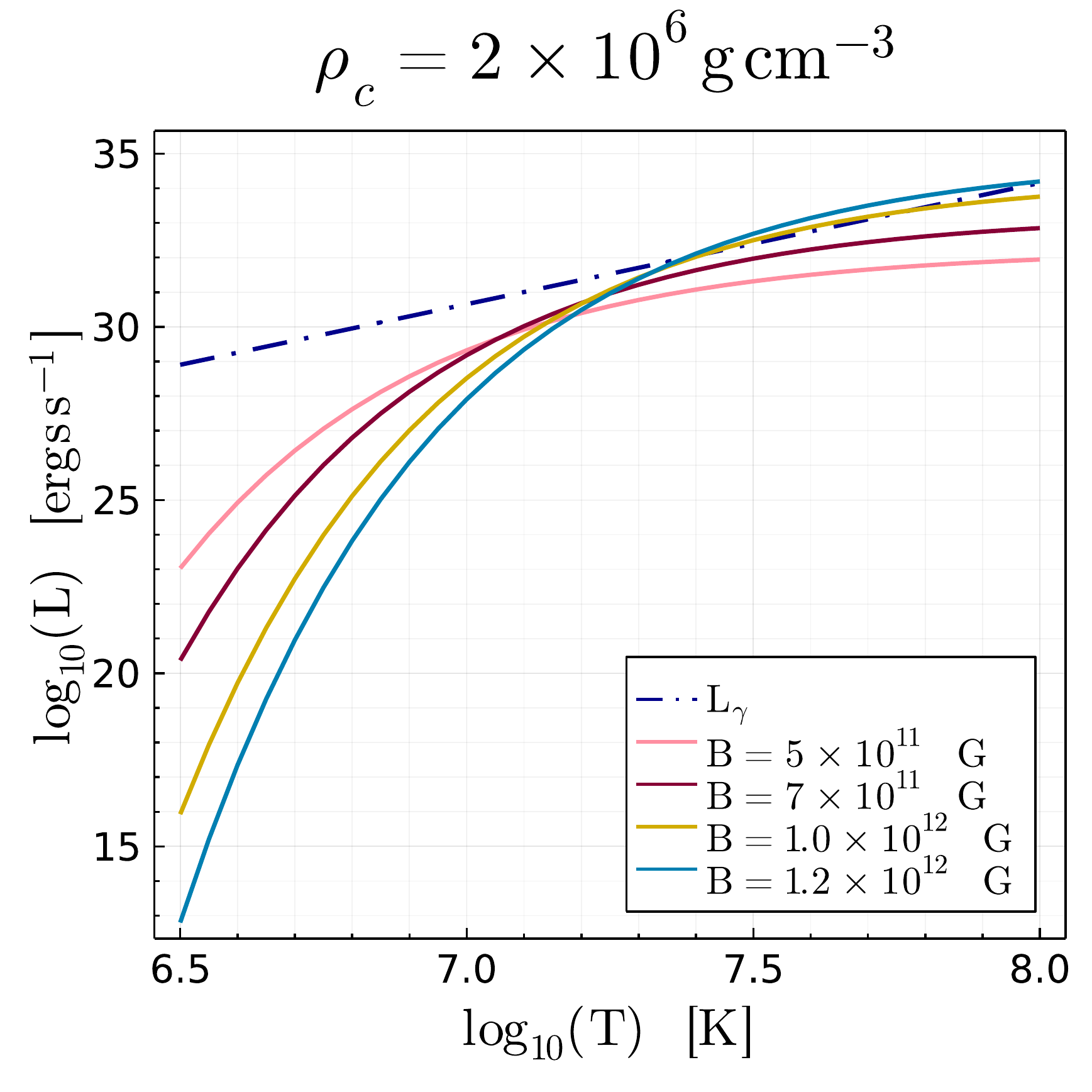}{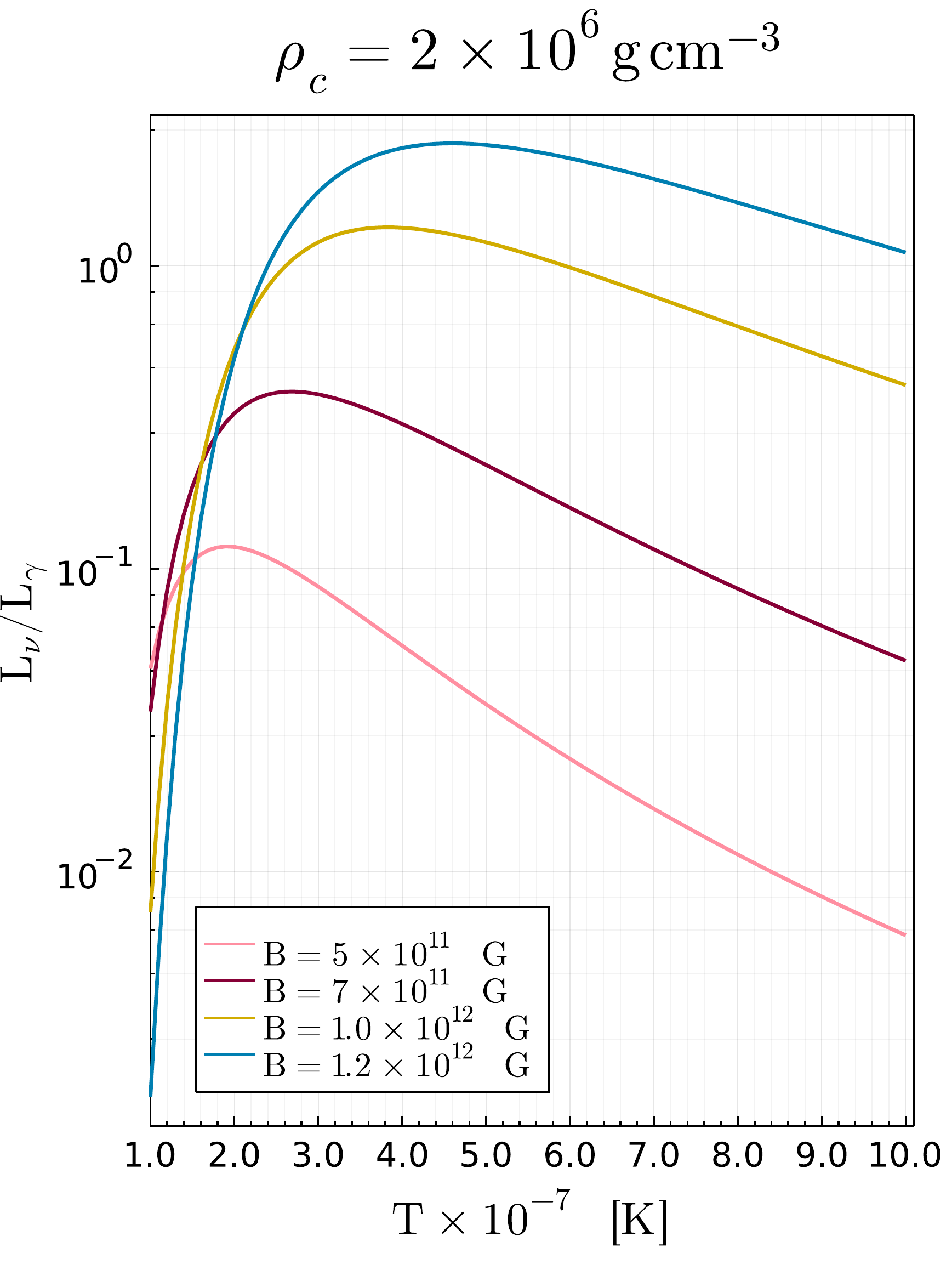}
    \caption{\textbf{Comparison between surface luminosity and neutrino luminosity.}  Neutrino emissivities are integrated over a white dwarf density profile with a central density $\rho_c=2\times 10^6 \, \rm g \, cm^{-3}$ for different values of the magnetic field.
    The ratio 
    shown in the right panel
    peaks at larger temperatures as larger values of the magnetic field are assumed.
    } 
    \label{fig:luminosities}
\end{figure}

In this appendix, we show how the neutrino pair synchrotron emissivity can be integrated over a WD profile to refine the rough results shown in the main text. Given the precision we aim for, we can assume the WD to be isothermal, while the density profile can be found assuming hydrostatic equilibrium and the electron gas equation of state to be a polytrope 
in the low-density limit. This choice is consistent with the non-relativistic approximations made throughout the paper, and the error introduced with respect to the exact equation of state (as well as relativistic and Coulomb corrections) is at most tens percent for a $0.6\, M_\odot$ WD.

Since we have shown that for $\omega_B\gtrsim T$ the weakly quantised degenerate limit is a good approximation, we can easily integrate Eq.~\eqref{QSynchWeakQuant} over a density profile, assuming a constant magnetic field over the entire core with a sharp drop near the surface, and compare its temperature evolution to \eqref{eq:mestellum},
\begin{equation}
		 L_\gamma = \vartheta L_{\odot}\frac{M}{M_\odot}  T_{7}^{3.5} \, .
	\end{equation}
	The results are shown in Fig.~\ref{fig:luminosities}. 
	We see that compared to the constraint reported in the main text, we get slightly more stringent bounds on the internal magnetic fields when considering a $\mathcal{O}(1)$ effect on the cooling ($1\times 10^{12} \, \rm G$, rather than $1.2\times 10^{12} \, \rm G$). With $B=5\times 10^{11}\,\rm G$ we find a $10\%$ effect.
	These results are easily explained by the dependence on the density of Eqs.~\eqref{VolumeAveragedQgamma} and~\eqref{QSynchWeakQuant}, since the first scales as $\rho$ and the second as $\rho^{1/3}$. Therefore, the assumption of a constant density made in the main text is conservative. Notice however that the neutrino pair synchrotron emission is strongly dependent on the magnetic field, so the bounds eventually obtained assuming a fixed density over the WD and a more realistic profile will differ by tens of percents.

\bibliography{whitedwarf}{}
\bibliographystyle{aasjournal}

\end{document}